\documentclass[aps,
notitlepage,twocolumn,showpacs,superscriptaddress,nofootinbib,preprintnumbers]{revtex4-1}  % for review and submission
\usepackage{graphicx}  % needed for figures
\usepackage{dcolumn}   % needed for some tables
\usepackage{bm}        % for math
\usepackage{amssymb}   % for math
\usepackage{amsmath}
\usepackage{slashed}
\usepackage{mathtools}
\usepackage[colorlinks=true,citecolor=cyan,urlcolor=teal]{hyperref}
\usepackage{cleveref}
\usepackage{tikz}
\usetikzlibrary{patterns}

\DeclarePairedDelimiter{\floor}{\lfloor}{\rfloor}
\usepackage{soul}

\allowdisplaybreaks

\newcommand{\cE}{\mathcal{E}}
\newcommand{\cI}{\mathcal{I}}
\newcommand{\cO}{\mathcal{O}}
\newcommand{\cM}{\mathcal{M}}

\begin{document}

\title{Leading-order gravitational radiation to all spin orders}

\author{Rafael Aoude}
\email[]{rafael.aoude@ed.ac.uk}
\affiliation{Centre for Cosmology, Particle Physics and Phenomenology (CP3),\\
Universit\'{e} catholique de Louvain, 1348 Louvain-la-Neuve, Belgium}
\affiliation{Higgs Centre for Theoretical Physics, School of Physics and Astronomy, \\
	The University of Edinburgh, Edinburgh EH9 3JZ, Scotland, UK}
\author{Kays Haddad}
\email[]{kays.haddad@physics.uu.se}
\affiliation{Department of Physics and Astronomy, Uppsala University, \\
Box 516, 75120 Uppsala, Sweden}
\affiliation{Nordita, Stockholm University and KTH Royal Institute of Technology, \\
Hannes Alfv\'{e}ns v\"{a}g 12, 10691 Stockholm, Sweden}
\author{Carlo Heissenberg}
\email[]{c.heissenberg@qmul.ac.uk}
\affiliation{School of Mathematical Sciences, Queen Mary University of London, Mile End Road, London, E1 4NS, United Kingdom}
\author{Andreas Helset}
\email[]{andreas.helset@cern.ch}
\affiliation{Walter Burke Institute for Theoretical Physics, California Institute of Technology,\\ Pasadena, CA 91125, USA}
\affiliation{Theoretical Physics Department, CERN, 1211 Geneva 23, Switzerland}

\date{\today}

\begin{abstract}
Starting with on-shell amplitudes compatible with the scattering of Kerr black holes, we produce the gravitational waveform and memory effect including spin at their leading post-Minkowskian orders to all orders in the spins of both scattering objects.
For the memory effect, we present results at next-to-leading order as well, finding a closed form for all spin orders when the spins are anti-aligned and equal in magnitude.
Considering instead generically oriented spins, we produce the next-to-leading-order memory to sixth order in spin.
Compton-amplitude contact terms up to sixth order in spin are included throughout our analysis.
\end{abstract}


\preprint{ IRMP-CP3-23-54}
\preprint{ UUITP-27/23}
\preprint{ CALT-TH-2023-040}
\preprint{ CERN-TH-2023-177}

\maketitle

%%%%%%%%%%%%%%%%%%%%%%%%%%%%%%%%%%%%%%%%%%%%%%%%%%%%%%%
%%%%%%% Begin Introduction %%%%%%%%%%%%%%%%%%%%%%%%%%%%
%%%%%%%%%%%%%%%%%%%%%%%%%%%%%%%%%%%%%%%%%%%%%%%%%%%%%%%

\section{Introduction}

The need for precision gravitational waveforms, which are crucial for detection and data analysis at LIGO, Virgo, Kagra, and associated experiments, has recently stimulated renewed effort in developing novel analytic techniques for calculating gravitational-wave observables relevant for binary encounters of compact objects.
In particular, scattering amplitudes provide compact, on-shell, and gauge invariant expressions that encode the dynamics of binary scatterings and their gravitational-wave emissions \cite{Bjerrum-Bohr:2018xdl,Kosower:2018adc,Bern:2019nnu,Bern:2019crd,Bern:2021dqo,Bern:2021yeh,Bern:2022jvn,Damgaard:2023vnx}; see refs.~\cite{Buonanno:2022pgc,Kosower:2022yvp} for recent reviews.
Organized in an expansion in the gravitational coupling, i.e., ~Newton's constant $G$, amplitudes are naturally suited for calculations in the weak-field, or post-Minkowskian (PM), regime.
Progress on the PM expansion has also come from worldline methods \cite{Kalin:2020mvi,Kalin:2020fhe,Liu:2021zxr,Dlapa:2021npj,Kalin:2022hph,Dlapa:2022lmu} and their close cousin the Worldline Quantum Field Theory (WQFT) \cite{Mogull:2020sak,Jakobsen:2021smu,Jakobsen:2021lvp,Jakobsen:2022fcj,Jakobsen:2023ndj,Jakobsen:2023hig}.

An important point concerns the inclusion of physical effects that go beyond the point-particle description of the scattering objects, notably those due to their tidal deformations \cite{Cheung:2020sdj,Haddad:2020que,Bern:2020uwk,Cheung:2020gbf,Aoude:2020ygw,AccettulliHuber:2020dal,Mougiakakos:2022sic,Heissenberg:2022tsn} and to their spins \cite{Arkani-Hamed:2017jhn,Vines:2017hyw,Guevara:2018wpp,Chung:2018kqs,Maybee:2019jus,Guevara:2019fsj,Arkani-Hamed:2019ymq,Johansson:2019dnu,Chung:2019duq,Damgaard:2019lfh,Bautista:2019evw,Aoude:2020onz,Chung:2020rrz,Bern:2020buy,Guevara:2020xjx,Kosmopoulos:2021zoq,Aoude:2021oqj,Chiodaroli:2021eug,Haddad:2021znf,Chen:2021kxt,Aoude:2022trd,Bern:2022kto,Alessio:2022kwv,FebresCordero:2022jts,Cangemi:2022abk,Alessio:2023kgf,Bautista:2023szu,Aoude:2023vdk,Heissenberg:2023uvo,Jakobsen:2023hig}, which can be introduced in the amplitude context by means of an effective-field-theory approach. 
A crucial conceptual issue consists in uniquely fixing the Wilson coefficients that are appropriate for describing a spinning black hole \cite{Chiodaroli:2021eug,Aoude:2022trd,Aoude:2022thd,Bautista:2021wfy,Cangemi:2022bew,Bautista:2022wjf,Bautista:2023szu,Aoude:2023vdk,Bjerrum-Bohr:2023iey}, and recent progress in this direction has been achieved by comparing amplitude calculations to fixed-background scattering described by the Teukolsky equation \cite{Bautista:2021wfy,Bautista:2022wjf}.

Several works have already endeavoured to produce state-of-the-art gravitational waveforms using scattering-amplitude or scattering-amplitude-inspired techniques.
In the former category, the Kosower-Maybee-O'Connell (KMOC) formalism \cite{Kosower:2018adc,Cristofoli:2021vyo} was recently employed in refs.~\cite{Brandhuber:2023hhy,Herderschee:2023fxh,Elkhidir:2023dco,Georgoudis:2023lgf,Caron-Huot:2023vxl} to connect the one-loop five-point amplitude with one graviton emission to the subleading PM waveform (see ref.~\cite{Bini:2023fiz} for a comparison with post-Newtonian results).
Also making use of the KMOC formalism, ref.~\cite{DeAngelis:2023lvf} produced the leading-order waveform for Kerr scattering up to fourth order in the spins of each black hole.

An alternative approach to the generation of waveforms is WQFT \cite{Mogull:2020sak,Jakobsen:2021smu,Jakobsen:2021lvp}, which was shown in ref.~\cite{Damgaard:2023vnx} to be equivalent to the extraction of observables through the KMOC formalism.
The applicability of this method to the generation of waveforms was demonstrated in ref.~\cite{Jakobsen:2021smu} through the derivation of the waveform without spin.
Ref.~\cite{Jakobsen:2021lvp} was then the first to include spin in the leading-order waveform, considering effects up to quadratic order in the spins of both black holes.

Yet another equivalent setup for extracting the leading order waveform employs the eikonal operator \cite{Cristofoli:2021jas,DiVecchia:2022piu}, in which graviton exchanges combine with coherent graviton emissions that build up gravitational waves (see ref.~\cite{DiVecchia:2023frv} for a review).

In this paper, we incorporate state-of-the-art knowledge about the amplitudes' description of Kerr black holes into the leading-PM gravitational waveform produced during a two-body encounter.
This observable is related by Fourier transform \cite{Jakobsen:2021lvp,Mougiakakos:2021ckm,Cristofoli:2021vyo} to (the factorizable portion of) the tree-level five-point amplitude describing the emission of a graviton from the scattering of two massive, spinning particles \cite{Bautista:2019tdr,Cristofoli:2021vyo,Bautista:2021inx,DeAngelis:2023lvf}.
We construct this amplitude recursively from the all-spin Kerr three-point amplitude \cite{Levi:2015msa,Vines:2017hyw,Guevara:2018wpp,Chung:2018kqs} and all-spin Kerr-compatible Compton amplitudes \cite{Aoude:2022trd,Haddad:2023ylx} (see also ref.~\cite{Bjerrum-Bohr:2023iey}).
Our Compton amplitude includes the contact terms up to sixth order in spin which are needed to match the black-hole-perturbation-theory (BHPT) description of (super-extremal) Kerr \cite{Bautista:2022wjf}.
Above sixth order in spin, the spinning objects described here deviate from Kerr only by contact terms in the Compton amplitude.
To accommodate for this discrepancy, we write the five-point amplitude and the waveform in a manner that automatically allows for the inclusion of higher-spin contact terms.

The waveform descending from the amplitude is presented to all spin orders in terms of two classes of arbitrary-tensor-rank integrals into impact-parameter space.
We explain the systematic evaluation of these integrals, and generate explicit results completing the waveform up to fifth order in spin in the ancillary files.
Describing the emitted graviton through spinor-helicity variables, we observe remarkable compactifications of the waveform stemming from the amplitude.
Illustrating this is a novel form of the leading-order waveform without spin; see also refs.~\cite{Kovacs:1978eu,DeAngelis:2023lvf,Jakobsen:2021smu}.

The low-frequency behavior of the spectral waveform, which translates to the one at early/late times via Fourier transform, is governed by soft theorems \cite{Sahoo:2018lxl,Laddha:2019yaj,Saha:2019tub,Sahoo:2021ctw}, which provide crucial non-perturbative cross-checks for PM calculations. Here, we leverage the universality of the leading soft theorem \cite{Weinberg:1964kqu,Weinberg:1965nx}---or memory effect in the time domain \cite{Zeldovich:1974gvh,Strominger:2014pwa}---which entirely fixes the leading soft behavior of the waveform sourced by the scattering objects in terms of their initial and final momenta, to calculate the memory to leading- and next-to-leading-PM orders.
At leading order we evaluate the memory to all spin orders and for generic orientations.
Making use of the all-spin 2PM amplitude derived in ref.~\cite{Aoude:2023vdk}, we produce the next-to-leading-PM memory for all spins when the spins are anti-aligned and equal in magnitude, and to sixth order for general configurations.

The paper is organized as follows.
In \cref{sec:FivePointAmplitude}, we construct the part of the all-spin five-point amplitude relevant to the waveform computation in \cref{sec:Waveform}.
The gravitational soft-theorem is applied to the extraction of the memory effect up to next-to-leading-PM order in \cref{sec:Memory}.
We conclude in Section~\ref{sec:Conclusions}.

\noindent\textbf{Note added:} On the day of submission of this paper, ref.~\cite{QMULSpinWaveform} appeared, which combines the integration method of ref.~\cite{DeAngelis:2023lvf} with the Compton amplitude of ref.~\cite{Bjerrum-Bohr:2023iey} to incorporate spin in the leading-order waveform.
The Compton amplitude employed in ref.~\cite{QMULSpinWaveform} exhibits spin-shift symmetry at fifth order in spin, which is in tension with the available BHPT data for super-extremal Kerr; see refs.~\cite{Aoude:2022trd,Bern:2022kto,Bautista:2022wjf} and \cref{sec:ComptonDetails}.

\section{Constructing the graviton-emission amplitude}
\label{sec:FivePointAmplitude}
%%%%%%%%%%%%%%%%%%%%%%%%%%%%%%%%%%%%%%%%%%%%%%%%%%%%%%%
\begin{figure}
    \centering
    \begin{tikzpicture}[scale=1, transform shape]
\usetikzlibrary{decorations.pathmorphing}
\tikzset{snake it/.style={decorate, decoration=snake}}
    \node (A2) at (2.5,-.75) {$+$};
    \node (A3) at (-.5,-.75) {$\uparrow q_{2}$};
    \node (A4) at (4.5,-.75) {$\downarrow q_{1}$};
    \node (A5) at (-1.0,.3) {$p_{1}\rightarrow$};
    \node (A6) at (-1.0,-1.8) {$p_{2}\rightarrow$};
    %\node (A7) at (0.2,0.7) {$k\nearrow$};
    \node (A7) at (0.7,-0.7) {$k\,$\rotatebox[origin=c]{-25}{$\rightarrow$}};
    %\node (A8) at (5.2,-2.2) {$k\searrow$};
    \node (A8) at (5.7,-0.8) {$k\,$\rotatebox[origin=c]{25}{$\rightarrow$}};
    \node (A9) at (1.0,.3) {$p_{1}^{\prime}\rightarrow$};
    \node (A10) at (1.0,-1.8) {$p_{2}^{\prime}\rightarrow$};
    \node (A11) at (4.0,.3) {$p_{1}\rightarrow$};
    \node (A12) at (4.0,-1.8) {$p_{2}\rightarrow$};
    \node (A13) at (6.0,.3) {$p_{1}^{\prime}\rightarrow$};
    \node (A14) at (6.0,-1.8) {$p_{2}^{\prime}\rightarrow$};
    \path [draw=blue] (-1.5,0) -- (1.5,0);
    \path [draw=red] (-1.5,-1.5) -- (1.5,-1.5);
    \path [draw=black, snake it] (0,-1.5) -- (0,0);
    %\path [draw=black, snake it] (0,0) -- (1.5,1.5);
    \path [draw=black, snake it] (0,0) -- (1.5,-0.7);
    \filldraw[fill=lightgray] (0,0) circle (6pt);
    \filldraw[fill=lightgray] (0,-1.5) circle (6pt);
    \path [draw=blue] (3.5,0) -- (6.5,0);
    \path [draw=red] (3.5,-1.5) -- (6.5,-1.5);
    \path [draw=black, snake it] (5.0,-1.5) -- (5.0,0);
    %\path [draw=black, snake it] (5.0,-1.5) -- (6.5,-3.0);
    \path [draw=black, snake it] (5.0,-1.5) -- (6.5,-0.8);
    \filldraw[fill=lightgray] (5.0,0) circle (6pt);
    \filldraw[fill=lightgray] (5.0,-1.5) circle (6pt);
\end{tikzpicture}
\caption{The cuts of the five-point amplitude relevant for the extraction of leading-order radiative observables. All shown momenta are taken on shell.}
\label{fig:FivePointCuts}
\end{figure}
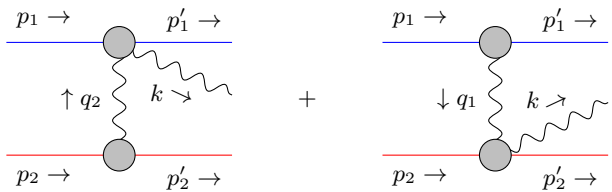
%%%%%%%%%%%%%%%%%%%%%%%%%%%%%%%%%%%%%%%%%%%%%%%%%%%%%%%

Classical radiative observables at leading order are related to the tree-level five-point amplitude with a graviton emitted from the scattering of two massive (spinning) particles.
More specifically, the relevant portion of the amplitude has non-vanishing residues when an internal graviton is taken on shell; see \cref{fig:FivePointCuts}.
Physically, the fact that only this portion of the amplitude is needed reflects the assumption that throughout the classical scattering the massive bodies are well separated.
As is evident from that figure, extracting the observables of interest thus requires the three-point and Compton amplitudes consistent with Kerr black holes.

The identification of classical spin effects in scattering amplitudes has been treated in numerous works \cite{Holstein:2008sx,Vaidya:2014kza,Guevara:2017csg,Guevara:2018wpp,Chung:2018kqs,Guevara:2019fsj,Maybee:2019jus,Damgaard:2019lfh,Aoude:2020onz,Bern:2020buy,Aoude:2021oqj,Cangemi:2022abk,Haddad:2023ylx}.
We will not review this material here, and will simply write spinning amplitudes directly in terms of the classical spin vector $S^{\mu}$ of an object of mass $m$ through the ring radius, $\mathfrak{a}^{\mu}=S^{\mu}/m$. 
This satisfies the covariant spin-supplementary condition $p\cdot\mathfrak{a}=0$, where $p^{\mu}$ is the classical momentum of the spinning object.

The three-point amplitude describing a Kerr black hole of momentum $p^{\mu}$, mass $m$, and ring radius $\mathfrak{a}^{\mu}$ emitting a helicity-$h$ graviton with momentum $q^{\mu}$ is \cite{Levi:2015msa,Vines:2017hyw,Guevara:2018wpp,Chung:2018kqs}
\begin{align}\label{eq:KerrThreePoint}
    \cM_{3}(-p,q^{h})=-\kappa \left[p\cdot\varepsilon_{h}(q)\right]^{2}\exp\left(hq\cdot\mathfrak{a}\right),
\end{align}
where $\kappa$ is related to Newton's constant $G$ through $\kappa=\sqrt{32\pi G}$.
The graviton polarization is $\varepsilon_{h}^{\mu\nu}(q)=\varepsilon_{h}^{\mu}(q)\varepsilon_{h}^{\nu}(q)$.\footnote{\label{footnote:WQFTPolNorm}We expect our results to be larger than those of refs.~\cite{Jakobsen:2021smu,Jakobsen:2021lvp} by a factor of $2$, due to differing conventions for graviton polarization tensors.}
Negative momentum arguments indicate incoming momenta.

A convenient writing of the Compton amplitude for the absorption of a graviton of momentum $q$ and emission of a (negative-helicity) graviton of momentum $k$ is
\begin{align}\label{eq:KerrCompton}
    &\cM_{4}(-p,k^{-},-q^{h}) \\
    &=\frac{\kappa^{2}}{4}\sum_{n=0}^{4}y_{h}^{4-n}\left(w_{h}\cdot\mathfrak{a}\right)^{n}M_{4}^{(n)}(-p,k^{-},-q^{h})\notag,
\end{align}
where $y_{h}\equiv2p\cdot w_{h}$ and 
\begin{align}
    w^{\mu}_{+}:=\frac{1}{2}\langle k|\sigma^{\mu}|q],
    \quad w^{\mu}_{-}:=\frac{1}{2m}\langle
    k|p\bar\sigma^{\mu}|q\rangle.
\end{align}
The form factors accompanying different powers of $w_{h}\cdot\mathfrak{a}$ are
\begin{align}
    M_{4}^{(n)}(-p,k^{-},-q^{h})&=e^{\left(-hq-k\right)\cdot\mathfrak{a}}F_{4}^{(n)}(-p,k^{-},-q^{h})\notag \\
    &\quad+C_{4}^{(n)}(-p,k^{-},-q^{h}).
\end{align}
The $F^{(n)}_{4}$ carry all the physical residues of the Compton amplitudes and the $C^{(n)}_{4}$ contain all information about contact deformations.
Our interest in leading-PM radiative observables for Kerr black holes requires that the $F_{4}^{(n)}$ are such that \cref{eq:KerrCompton} factorizes to \cref{eq:KerrThreePoint} on physical residues \cite{Aoude:2022trd}, and that the $C_{4}^{(n)}$ contain contact terms whose coefficients depend only on the mass (and not on $G$) \cite{Haddad:2023ylx}. 
We have left the dependence of the form factors on the ring radius implicit and relegate their explicit expressions to the appendices; see \cref{eq:FactorizableFFSH1,eq:FactorizableFFSH2,eq:FactorizableFFOH1,eq:FactorizableFFOH2,eq:ContactFFSH,eq:ContactFFOH0,eq:ContactFFOH1,eq:ContactFFOH2,eq:ContactFFOH3,eq:ContactFFOH4}.

The five-point amplitude is constructed from these lower-point amplitudes by demanding that it factorizes correctly on the physical graviton poles; see fig.~\ref{fig:FivePointCuts}. The momenta obey the momentum-conservation constraints
\begin{align}
    p_{i}-p_{i}^{\prime}=q_{i},\quad q_{1}+q_{2}=k.
\end{align}
We abbreviate the cut part of the five-point amplitude as $\cM^{\rm cut}_{5}(k^-)\equiv\cM^{\rm cut}_5(-p_{1},-p_{2},p_{1}^{\prime},p_{2}^{\prime},k^-)$.  Putting everything together, the cut part of the five-point amplitude relevant to Kerr observables at leading PM order is
\begin{align}\label{eq:FivePointAmplitude}
    &\cM^{\rm cut}_{5}(k^{-})=-\frac{\kappa^{3}}{8q^{2}_{2}}q_{2\mu}q_{2\nu} \\
    &\times\sum_{h=\pm}\sum_{n=0}^{4}\frac{r_{(1),n}^{h,\mu\nu}}{2^{n}}e^{-hq_{2}\cdot\mathfrak{a}_{2}}M_{4}^{(n)}(-p_{1},k^{-},-q^{h}_{2})+(1\leftrightarrow2),\notag
\end{align}
valid to all spin orders.
The helicity weight of the emitted graviton is carried by the $r_{(i),n}^{h,\mu\nu}$ (recall that $h$ here is the helicity of the cut graviton, not the emitted one), which are defined in \cref{eq:HelicityVectors}.
This abbreviation of the amplitude is useful for making explicit the powers of $q_{2}^{\mu}$ while hiding what is not needed to perform the waveform integration.
However, we highlight that the $r_{(i),n}^{h,\mu\nu}$ are $\cO(|k^{\mu}|^{2})$ and $\cO(\mathfrak{a}_{i}^{n})$.
The former of these will affect the integration to the time domain from frequency space.

At this point, let us make a remark on notation.
Throughout the remainder of the paper, we will use the subscripts in parentheses $(1)$ or $(2)$ to denote quantities relevant to the amplitude on the $q_{2}^{2}$ or $q_{1}^{2}$ pole, respectively.
An object indexed in this way has only the labels in parentheses swapped under the relabelling $(1\leftrightarrow2)$, so that, for example, $r^{h,\mu\nu}_{(1),1}\rightarrow r^{h,\mu\nu}_{(2),1}$, while $p_{1}^{\mu}\rightarrow p_{2}^{\mu}$.
Generally,
\begin{equation}
\begin{aligned}\label{eq:RelabellingNotation}
    X_{1}&\overset{(1\leftrightarrow2)}{\leftrightarrow}X_{2}, \\
    Y_{(1),L}&\overset{(1\leftrightarrow2)}{\leftrightarrow} Y_{(2),L}=\left.Y_{(1),L}\right|_{Z_{(1),J}\leftrightarrow Z_{(2),J},X_{1}\leftrightarrow X_{2}},
\end{aligned}
\end{equation}
where $L$ and $J$ are arbitrary (multi-)indices which remain unchanged under the relabelling.
The second relabelling is applied recursively at every level of an expression.

The five-point amplitude for the other graviton helicity is related to \cref{eq:FivePointAmplitude} through
\begin{align}
	\cM^{\rm cut}_{5}(k^{+})=\left[\cM^{\rm cut}_{5}(-k^{-})\right]^{*}_{q_{i}\rightarrow-q_{i}},
\end{align}
where the asterisk represents complex conjugation.
The effect of conjugation is simply swapping the angle and square massless spinors.

By construction, the cut part of the five-point amplitude in \cref{eq:FivePointAmplitude} gives the correct factorization when an exchanged graviton goes on shell. 
This is achieved by writing the amplitude in a form with spurious poles using the identity $\frac{1}{q_1^2 q_2^2} = - \frac{1}{2(k\cdot q_1) q_1^2} - \frac{1}{2(k\cdot q_2) q_2^2}$ \cite{Bautista:2019tdr}, which ensures that the physical graviton poles are not overlapping and the cut part of the amplitude can be constructed by gluing the lower-point amplitudes.
However, it does not guarantee that the spurious poles cancel after the gluing.
In fact, the freedom of rewriting the lower-point amplitudes using on-shell conditions and momentum conservation allows for various representations of the cut part of the amplitude which differ by terms that vanish when internal gravitons are cut.
In particular, the difference between \cref{eq:FivePointAmplitude} and the complete five-point amplitude with no unphysical poles has no pole when internal gravitons go on shell.
Luckily, these complications are irrelevant for extracting the waveform from the amplitude because terms with a spurious pole and no physical graviton pole do not contribute.
This is a consequence of the classical limit, and manifests as the tracelessness of the integrals in the next section; see \cref{sec:Integrals}.

\section{All-spin waveform at leading-PM order}\label{sec:Waveform}

With the cut part of the five-point amplitude in hand, we move now to the extraction of observables.
The waveform in the time domain is given by the expectation value of the metric perturbation \cite{Cristofoli:2021vyo,Kosower:2022yvp,Brandhuber:2023hhy,Herderschee:2023fxh,Elkhidir:2023dco,Georgoudis:2023lgf}:
\begin{align}
\begin{aligned}
    g_{\mu\nu}(x)-\eta_{\mu\nu}&=
    \kappa\langle h_{\mu\nu}(x) \rangle \\
    &= \int_{-\infty}^{\infty} \hat d \omega\, e^{-i\omega u} \frac{f_{\mu\nu}(\omega,\hat{\boldsymbol{x}})}{|\boldsymbol{x}|}.\label{eq:WaveformDefinition}
\end{aligned}
\end{align}
Here, $\omega$ is the frequency of the emitted gravitational wave, $x^{\mu}=(x^{0},\boldsymbol{x})$ is the position of the observer located a large distance from the scattering event, and $u=x^{0}-|\boldsymbol{x}|$ is the retarded time.
The spatial unit vector in the direction of the observer is $\hat{\boldsymbol{x}}=\boldsymbol{x}/|\boldsymbol{x}|$.
The spectral waveform $f_{\mu\nu}(\omega,\hat{\boldsymbol{x}})$ is written at leading order in terms of the tree-level five-point amplitude as
\begin{align}\label{eq:SpectralWaveform}
    &f_{\mu\nu}(\omega,\hat{\boldsymbol{x}})=\frac{\kappa}{4\pi}\sum_{h}\varepsilon_{\mu}^{(h)*}\varepsilon_{\nu}^{(h)*}\left.\int_{q_{1},q_{2}}\mu(k)\cM^{\rm cut}_{5}(k^{h})\right|_{k=\omega\rho},
\end{align}
where $\rho^{\mu}=(1,\hat{\boldsymbol{x}})$, $\int_{q_{1},q_{2}}=\int \hat{d}^{D}q_{1}\hat{d}^{D}q_{2}$, and
\begin{align}
	\mu(k)  =& \frac{1}{4}\hat \delta(p_1 \cdot q_1) \hat \delta(p_2 \cdot q_2) e^{i(q_1 \cdot b_1 + q_2 \cdot b_2)} \hat \delta^{D}(q_1 + q_2 - k) \,. \nonumber
\end{align}
Factors of $2\pi$ have been absorbed into the notation as $\hat\delta(x)\equiv 2\pi\delta(x)$ and $\hat{d}x\equiv dx/(2\pi)$~\cite{Kosower:2018adc}, and we work in four dimensions, $D=4$.

The sum over helicities in \cref{eq:SpectralWaveform} can be dropped by projecting onto a graviton of a fixed helicity.
We can do so without losing any information about the waveform since the projection onto a graviton of the opposite helicity will be given by complex conjugation of our result.
In the following, we write $\varepsilon_{\mu}^{-}\varepsilon_{\nu}^{-}\langle h^{\mu\nu}(x)\rangle=h(x)=h_{+}(x)+ih_{\times}(x)$, where the subscripts refer to the ``plus" and ``cross" polarizations of the gravitational wave.

Inserting \cref{eq:FivePointAmplitude} into \cref{eq:WaveformDefinition}, we proceed to integrate following ref.~\cite{Jakobsen:2021lvp}.
Specifically, on the part of the cut amplitude capturing the $q_{2}^{2}$ residue, it is advantageous to use the delta functions to integrate over $\hat{d}^{4}q_{1}$ and $\hat{d}\omega$ first; on the other part of the amplitude, one instead integrates over $\hat{d}^{4}q_{2}$ before performing the $\hat{d}\omega$ integral.
Indeed, this procedure remains simple for the infinite-spin amplitude in \cref{eq:FivePointAmplitude}.
Splitting the waveform into a part without and with Compton-amplitude contact terms,
\begin{align}\label{eq:WaveformDecomposed}
    \kappa h(x)=-\frac{\pi G^{2}}{|\boldsymbol{x}|m_{1}m_{2}}\left[h_{f}(x)+h_{c}(x)\right],
\end{align}
and writing $p_i^{\mu} = m_i v_i^{\mu}$ and $\gamma = v_1 \cdot v_2$,\footnote{In previous works using Heavy Particle Effective Theory (HPET) \cite{Damgaard:2019lfh,Aoude:2020onz,Haddad:2020tvs,Haddad:2020que,Aoude:2020ygw,Haddad:2021znf,Aoude:2022trd,Aoude:2022thd,Haddad:2023ylx,Aoude:2023vdk}, the symbol $\omega = v_1 \cdot v_2$ was used as an homage to the literature on Heavy Quark Effective Theory \cite{Georgi:1990um,Isgur:1989vq,Isgur:1990yhj,Manohar:2000dt}. In this paper, $\omega$ is reserved for the frequency of the gravitational wave, so we revert to the notation here.} the former is
\begin{align}\label{eq:WaveformImplicitIntegral}
    &h_{f}(x)=\frac{1}{(p_{1}\cdot\rho)^{2}} \\
    &\times\int_{q_{2}}\hat \delta(v_2 \cdot q_2)\frac{q_{2\mu_{1}}q_{2\mu_{2}}}{q^{2}_{2}(q_{2}\cdot\rho)(v_{1}\cdot q_{2})}\left[e^{iq_2 \cdot b_{(1),-}}\tilde{r}_{(1),0}^{-,\mu_{1}\mu_{2}}\right.\notag \\
    &\left.+e^{iq_2 \cdot b_{(1),+}}\sum_{s=0}^{\infty}q_{2\mu_{3}}\dots q_{2\mu_{s+2}}\frac{1}{s!}\mathcal{L}_{(1),s}^{\mu_{1}\dots\mu_{s+2}}\right]+(1\leftrightarrow2).\notag
\end{align}
In a similar vein to ref.~\cite{Jakobsen:2021lvp}, we have defined
\begin{align}
	u^{\pm}_{(1),1}&=\frac{\rho\cdot[(x-i\mathfrak{a}_{1})-b_{1}]}{v_{1}\cdot\rho},\label{eq:Waveformu11} \\
	u^{\pm}_{(1),2}&=\frac{\rho\cdot[(x-i\mathfrak{a}_{1})-b_{2} \mp i (\mathfrak{a}_1 + \mathfrak{a}_2)]}{v_{2}\cdot\rho},\label{eq:Waveformu12} \\
    b^{\mu}_{(1),\pm}&=b_{2}^{\mu}-b_{1}^{\mu}+u^{\pm}_{(1),2}v_{2}^{\mu}-u^{\pm}_{(1),1}v_{1}^{\mu}\pm i(\mathfrak{a}_{1}^{\mu}+\mathfrak{a}_{2}^{\mu}).\label{eq:ShiftedIP}
\end{align}
These variables expose a Newman-Janis-like shift of the position coordinates by the spin vector \cite{Newman:1965tw}.
Unlike the Newman-Janis shift, however, the amplitude contains spin dependence which does not readily admit this interpretation, such as in the $\tilde{r}_{(i),n}^{h,\mu\nu}$ and $\mathcal{L}_{(i),s}^{\mu_{1}\dots\mu_{s+2}}$.
Nevertheless, the presence of this shift is computationally convenient, as it implies that the highest tensor rank needed to obtain the $\cO(\mathfrak{a}_{1}^{n_{1}}\mathfrak{a}_{2}^{n_{2}})$ part of the waveform is $\max(n_{1},n_{2})+2$ instead of $n_{1}+n_{2}+2$.  
The tensors $\tilde{r}_{(i),n}^{h,\mu\nu}$ are functions only of $\rho^{\mu}$ defined through $r_{(i),n}^{h,\mu\nu}=\omega^{2}\tilde{r}_{(i),n}^{h,\mu\nu}$.
Finally, $\mathcal{L}_{(i),s}^{\mu_{1}\dots\mu_{s+2}}$ is given in \cref{eq:FormFactorNoq}.
This tensor is a complicated polynomial of degree $s$ in the spin of particle $i$, which, importantly, contains no dependence on the final variable of integration, and can therefore be removed from the integral.
Then, defining
\begin{align}\label{eq:WaveformFactorizableIntegral}
    \mathcal{I}^{\mu_{1}\dots\mu_{n}}_{(1)}(b)\equiv\int_{q_{2}}\hat \delta(v_2 \cdot q_2)\frac{q_{2}^{\mu_{1}}\dots q_{2}^{\mu_{n}}e^{iq_{2}\cdot b}}{q^{2}_{2}(q_{2}\cdot\rho)(v_{1}\cdot q_{2})},
\end{align}
the part of the waveform free from Compton-amplitude contact terms is
\begin{align}\label{eq:WaveformImplicit}
    &h_{f}(x)=\frac{1}{(p_{1}\cdot\rho)^{2}}\left[\tilde{r}_{(1),0}^{-,\mu_{1}\mu_{2}}\mathcal{I}_{(1),\mu_{1}\mu_{2}}(b_{(1),-})\right.\notag \\
    &\left.+\sum_{s=0}^{\infty}\frac{1}{s!}\mathcal{L}_{(1),s}^{\mu_{1}\dots\mu_{s+2}}\mathcal{I}_{(1),\mu_{1}\dots\mu_{s+2}}(b_{(1),+})\right]+(1\leftrightarrow2).
\end{align}
We are left now with the evaluation of the arbitrary-rank integral in \cref{eq:WaveformFactorizableIntegral}.
The variables defined in \cref{eq:Waveformu11,eq:Waveformu12,eq:ShiftedIP} produce $\rho\cdot b_{(i),\pm}=0$, which means that the integrals appearing in \cref{eq:WaveformImplicit} are identical in structure to the ones in ref.~\cite{Jakobsen:2021lvp}, justifying the use of the rank-2 integral evaluated there.
Higher-rank integrals can be generated by differentiation, keeping in mind that the result must remain orthogonal to $v_{2}^{\mu}$:
\begin{align}
    \mathcal{I}^{\mu\nu\sigma_{1}\dots\sigma_{n}}_{(1)}(b_{(1),\pm})&=\left(\prod_{i=1}^{n}\frac{-i\partial}{\partial \boldsymbol{b}_{(1),\pm,\sigma_{i}}}\right)\mathcal{I}^{\mu\nu}_{(1)}(b_{(1),\pm}),
\end{align}
for $\boldsymbol{b}_{(1),\pm}^{\sigma}\equiv\left(\delta^{\sigma}_{\tau}-v_{2}^{\sigma}v_{2\tau}\right)b_{(1),\pm}^{\tau}$.
More details can be found in \cref{sec:Integrals}.

As an illustration of \cref{eq:WaveformImplicit}, consider the spinless part of the waveform.
When setting the spin to 0, $b_{(i),+}^{\mu}|_{\mathfrak{a}_{i}=0}=b_{(i),-}^{\mu}|_{\mathfrak{a}_{i}=0}$ and $b_{(1),\pm}^{\mu}|_{\mathfrak{a}_{i}=0}=-b_{(2),\pm}^{\mu}|_{\mathfrak{a}_{i}=0}\equiv b_{0}^{\mu}$.
Then, since the rank-2 integral is even (see \cref{eq:wfInt}), the waveform is
\begin{align}\label{eq:WaveformSpinless}
    h_{f}(x)|_{\mathfrak{a}_{i}=0}&=\sum_{i=1}^{2}\frac{\tilde{r}_{(i),0}^{-,\mu\nu}+\tilde{r}_{(i),0}^{+,\mu\nu}}{(p_{i}\cdot\rho)^{2}}\mathcal{I}_{(i),\mu\nu}(b_{0}),
\end{align}
in agreement with refs.~\cite{Kovacs:1978eu,Jakobsen:2021smu,Jakobsen:2021lvp,DeAngelis:2023lvf}.
Much like for scattering amplitudes, the incorporation of spinor-helicity variables greatly compactifies the form of the waveform, eliminating gauge redundancies associated with the polarization of the emitted graviton; cf.~the compact eq.~(32) of ref.~\cite{DeAngelis:2023lvf}, which expresses this same result using polarization tensors.

The extraction of the contributions to the waveform originating from the Compton-amplitude contact terms is slightly different from above because of the simpler pole structure which enters the Fourier transforms.
Explicitly, the contact-term contribution to the waveform is
\begin{align}\label{eq:WaveformContacts}
    &h_{c}(x)=\frac{32m_{1}v_{1}^{\mu_{1}}v_{1}^{\mu_{2}}v_{1}^{\mu_{3}}}{(v_1 \cdot \rho)^{3}}\left[C_{4}^{(5,1),\mu_{4}\mu_{5}}(\mathfrak{a}_{1})\mathcal{J}_{(1),\mu_{1}\dots\mu_{5}}(b_{(1)})\right.\notag \\
    &\left.+\mathcal{J}_{(1),\mu_{1}\dots\mu_{6}}(b_{(1)})\sum_{i=1}^{3}C_{4}^{(6,i),\mu_{4}\mu_{5}\mu_{6}}(\mathfrak{a}_{1})\right]+(1\leftrightarrow2),
\end{align}
where we've used $C_{4}^{(n)}(-p_{1},k^{-},-q_{2}^{-})=0$ and repackaged the remaining contact terms in the $C_{4}^{(i,j)}$.
The explicit forms for these can be found in \cref{eq:RepackagedContact1,eq:RepackagedContact2,eq:RepackagedContact3,eq:RepackagedContact4}.
Organized like this, the $C_{4}^{(i,j)}(\mathfrak{a}_{1})$ are all free of the variable of integration, so we have removed them from the integrals
\begin{align}\label{eq:WaveformContactIntegral}
    \mathcal{J}^{\mu_{1}\dots\mu_{n}}_{(1)}(b_{(1)})&=\int_{q_{2}}\hat{\delta}(v_{2}\cdot q_{2})\frac{e^{iq_{2}\cdot b_{(1)}}}{q_{2}^{2}}q_{2}^{\mu_{1}}\dots q_{2}^{\mu_{n}}.
\end{align}
The impact parameter in this context is $b_{(1)}^{\mu}=b_{(1),+}^{\mu}|_{\mathfrak{a}_{1}\rightarrow0}$.
This integral for $n=1$ has also been evaluated in ref.~\cite{Jakobsen:2021lvp}, with higher-rank integrals being generated by differentiating with respect to $\boldsymbol{b}^{\mu}_{(1)}$.
See \cref{sec:Integrals} for more details.

Two final remarks about $h_{c}(x)$ are in order.
First, note that all dependence on $\mathfrak{a}_{2}^{\mu}$ in \cref{eq:WaveformContacts} ($\mathfrak{a}_{1}^{\mu}$ in the relabelled part) is encapsulated in the impact parameter.
Consequently, \cref{eq:WaveformContacts} encodes the contributions from the $\cO(\mathfrak{a}^{5,6})$ coefficients to \textit{all} spin orders.
Second, the inclusion of higher-spin contact terms is nearly automatic: contact terms at $\cO(\mathfrak{a}^{s})$ enter the square brackets of \cref{eq:WaveformContacts} through
\begin{align}
    v_{1\mu_{1}}\mathcal{J}_{(1),\mu_{2}\dots\mu_{s+1}}(b_{(1)})C_{4}^{(s),\mu_{4}\dots\mu_{s+1}}(\mathfrak{a}_{1}).
\end{align}
All that must be specified are the contact terms one wishes to include in the $C_{4}^{(s)}$.

With that, we have produced the leading-order Kerr-compatible waveform to all spin orders, including BHPT-matching contact terms up to sixth order in spin.
The waveform expanded up to fifth order in spin is provided in the ancillary files.
Our results agree with ref.~\cite{Jakobsen:2021lvp} up to second order in the spin, and with ref.~\cite{DeAngelis:2023lvf} up to fourth order in the spin.

Further checks of our results come from the expansion of the waveform in large $|u|$ (frequency space, small $\omega$).
In the next section we will consider the memory effect and its connection to the leading classical soft theorem.
The subleading classical soft theorem predicts instead the $1/|u|$ (frequency space, $\log\omega$) tail term \cite{Saha:2019tub}, which is spin-independent to this order in $G$; we have verified its agreement with the spinless part of the waveform given in \cref{eq:WaveformSpinless}.
To this order in $G$, the next order in the soft expansion features a $1/u^2$ (frequency space, $\omega\log\omega$) term whose expression was predicted in ref.~\cite{Ghosh:2021bam} and contains both spin-independent and linear-in-spin contributions.\footnote{We thank Biswajit Sahoo for bringing this to our attention.} 
We have verified that our waveform agrees with that prediction---the spinless and linear-in-spin contributions to the waveform exactly match the $1/u^2$ soft term given in ref.~\cite{Ghosh:2021bam}, while higher-spin contributions decay faster than $1/u^2$ for large $|u|$.

\section{Gravitational memory effect}
\label{sec:Memory}

Rather than requiring the full five-point amplitude, or even its cut part used above, the gravitational memory effect is related to the limit of the five-point amplitude as the emitted graviton is soft \cite{Strominger:2014pwa}.
Combining this with existing high-spin, two-to-two amplitudes compatible with Kerr scattering up to 2PM order puts the next-to-leading-order (NLO) memory effect including high spin orders within reach \cite{Guevara:2019fsj,Aoude:2023vdk} (see also refs.~\cite{Chen:2021kxt,Bern:2022kto,Bautista:2023szu} for high-but-finite-spin scattering amplitudes at 2PM order).
In this section we present the gravitational memory effect at leading order to all spin orders and for generic spin orientations, before computing the next-to-leading-order memory effect to all spin orders for anti-aligned spins and to sixth order for generic orientations.

\subsection{Leading order}

At leading order in Newton's constant, the memory effect is expressed simply in terms of the soft limit of the tree-level two-to-two amplitude as \cite{Strominger:2014pwa,Herderschee:2023fxh}
\begin{align}\label{eq:LOMemoryST}
    &\left.\Delta(h_{+}^{\infty}+ih_{\times}^{\infty})\right|^{\rm LO}=-\frac{i\kappa}{32\pi|\boldsymbol{x}|} \\
    &\qquad\times\int\hat{d}^{4}q\,\hat{\delta}(p_{1}\cdot q)\hat{\delta}(p_{2}\cdot q)e^{iq\cdot b}\varepsilon^{\mu}_{-}\varepsilon^{\nu}_{-}\mathcal{S}(\rho,q)_{\mu\nu}\cM_{t}(q).\notag
\end{align}
Here, $\mathcal{S}_{\mu\nu}(\rho,q)=\omega\mathcal{S}_{\mu\nu}(k,q)$ is the soft factor multiplied by the frequency of the soft graviton, and $\cM_{t}(q)$ is the $t$-channel graviton-exchange amplitude.
Orienting the transfer momentum such that $q=p_{1}-p_{1}^{\prime}=p_{2}^{\prime}-p_{2}$, these take the forms \cite{Weinberg:1965nx,Guevara:2019fsj}
\begin{align}
    \mathcal{S}_{\mu\nu}(k,q)&=\sum_{i=1}^{2}\frac{p_{i}^{\mu}p_{i}^{\nu}}{p_{i}\cdot k}-\sum_{i=1}^{2}\frac{p_{i}^{\prime\mu}p_{i}^{\prime\nu}}{p_{i}^{\prime}\cdot k},\label{eq:LeadingSoftFactor} \\
    \cM_{t}(q)&=-\frac{\kappa^{2}m_{1}^{2}m_{2}^{2}}{4q^{2}}\sum_{\pm}\left(\gamma\pm\sqrt{\gamma^{2}-1}\right)^{2}\label{eq:1PMAmplitude} \\
    &\quad\times\exp\left[\pm\frac{i\epsilon_{\mu\nu\alpha\beta}v_{1}^{\mu}v_{2}^{\nu}q^{\alpha}}{\sqrt{\gamma^{2}-1}}\left(\mathfrak{a}_{1}^{\beta}+\mathfrak{a}_{2}^{\beta}\right)\right].\notag
\end{align}
The soft factor admits an $\hbar$ expansion when scaling $k,q\sim\hbar$, which we can write as
\begin{align}
    \mathcal{S}_{\mu\nu}(k,q)=\mathcal{S}^{(0)}_{\mu\nu;\rho}(k)q^{\rho}+\mathcal{S}^{(1)}_{\mu\nu;\rho\tau}(k)q^{\rho}q^{\tau}+\cO(\hbar^{2}),\label{eq:SoftFactorExpanded}
\end{align}
with $\mathcal{S}^{(i)}_{\mu\nu;\rho_{1}\dots\rho_{i+1}}(k)\sim\hbar^{-1}$.
The classically-relevant portion of \cref{eq:LOMemoryST} only needs the leading term in \cref{eq:SoftFactorExpanded}.

The fact that the spin dependence of the amplitude in \cref{eq:1PMAmplitude} is contained entirely in an exponential means that the evaluation of the gravitational memory effect for all spins at leading order is nearly identical to the scalar case \cite{Alessio:2022kwv}.
Defining
\begin{align}
    b^{\mu}_{\pm}\equiv b^{\mu}\pm \frac{\epsilon_{\mu\nu\alpha\beta}v_{1}^{\nu}v_{2}^{\alpha}\left(\mathfrak{a}_{1}^{\beta}+\mathfrak{a}_{2}^{\beta}\right)}{\sqrt{\gamma^{2}-1}},
\end{align}
the memory effect is
\begin{align}
    &\left.\Delta(h_{+}^{\infty}+ih_{\times}^{\infty})\right|^{\rm LO}=\frac{i\kappa^{3}m_{1}^{2}m_{2}^{2}}{128\pi|\boldsymbol{x}|}\sum_{\pm}\left(\gamma\pm\sqrt{\gamma^{2}-1}\right)^{2}\notag \\
    &\times\varepsilon^{\mu}_{-}\varepsilon^{\nu}_{-}\mathcal{S}^{(0)}_{\mu\nu;\tau}(\rho)\int\hat{d}^{4}q\,\hat{\delta}(p_{1}\cdot q)\hat{\delta}(p_{2}\cdot q)\frac{e^{iq\cdot b_{\pm}}}{q^{2}}q^{\tau}.\label{eq:MemorySoftExplicit}
\end{align}
From this expression, it is immediate to explain an observation made in  ref.~\cite{Jakobsen:2021lvp} up to quadratic order in spin and to see that it extends to all spin orders: for the anti-aligned-spin setup, $\mathfrak{a}_{1}^{\mu}=-\mathfrak{a}_{2}^{\mu}$, the leading-order memory effect for two scattering Kerr black holes is equivalent to that for Schwarzschild black holes.
This is because the shifted impact parameters are identical to the unshifted impact parameter in this configuration.

We recognize in \cref{eq:MemorySoftExplicit} the expression for the leading-order classical impulse, $\mathcal{Q}_{\rm 1,1PM}^{\mu}=p_{1}^{\prime\mu}-p_{1}^{\mu}$, experienced by particle 1 in the scattering \cite{Kosower:2018adc}:
\begin{align}
    \mathcal{Q}_{1,{\rm 1PM}}^{\tau}&=-\frac{i}{4}\int\hat{d}^{4}q\hat{\delta}(p_{1}\cdot q)\hat{\delta}(p_{2}\cdot q)q^{\tau}e^{ib\cdot q}\cM_{t}(q) \\
    &=\frac{\kappa^{2}m_{1}m_{2}}{32\pi\sqrt{\gamma^{2}-1}}\sum_{\pm}\left(\gamma\pm\sqrt{\gamma^{2}-1}\right)^{2}\frac{b_{\pm}^{\tau}}{b_{\pm}^{2}}.\notag
\end{align}
In terms of the impulse, the leading-order, all-spin memory effect is
\begin{align}
    &\left.\Delta(h_{+}^{\infty}+ih_{\times}^{\infty})\right|^{\rm LO}=\frac{\kappa\varepsilon^{\mu}_{-}\varepsilon^{\nu}_{-}}{8\pi|\boldsymbol{x}|}\mathcal{S}^{(0)}_{\mu\nu;\tau}(\rho)\mathcal{Q}^{\tau}_{1,{\rm 1PM}},\label{eq:LOMemoryResult}
\end{align}
with generic spin orientations.
Eq.~\eqref{eq:LOMemoryResult} thus agrees with the leading-PM expansion of the leading classical soft theorem \cite{Saha:2019tub,Sahoo:2021ctw,DiVecchia:2022owy,DiVecchia:2022nna}, which fixes the memory effect in the time domain, equivalently the $1/\omega$ terms in frequency domain, in terms of the initial and final momenta of the scattering.
We have also checked that the spinless contribution is in agreement with ref.~\cite{Herderschee:2023fxh}.
Specializing to the aligned-spin case and expanding to quadratic order in spin, we find agreement with the result in ref.~\cite{Jakobsen:2021lvp}, taking into account the factor of 2 mentioned in \cref{footnote:WQFTPolNorm}.

\subsection{Next-to-leading order}

To subleading PM order, following ref.~\cite{Caron-Huot:2023vxl}, we modify the integrand of \cref{eq:LOMemoryST} to contain the classical part of the 2PM instead of the tree-level amplitude and include a two-massive-particle cut contribution that arises from the KMOC formalism.\footnote{We thank Zvi Bern for discussions on this point.}
Additional classical cuts involving intermediate on-shell massless and massive particle lines are subleading in the soft limit \cite{Brandhuber:2023hhy,Herderschee:2023fxh,Elkhidir:2023dco,Georgoudis:2023lgf}.
The next-to-leading-order memory effect is thus given by
\begin{widetext}
\begin{align}\label{eq:NLOMemoryST}
    &\left.\Delta(h_{+}^{\infty}+ih_{\times}^{\infty})\right|^{\rm NLO}=-\frac{i\kappa}{32\pi|\boldsymbol{x}|}\int\hat{d}^{4}q\,\hat{\delta}\left(p_{1}\cdot q-\frac{q^{2}}{2}\right)\hat{\delta}\left(p_{2}\cdot q+\frac{q^{2}}{2}\right)e^{iq\cdot b}\varepsilon^{\mu}_{-}\varepsilon^{\nu}_{-}\notag \\
    &\times\left\{\mathcal{S}(\rho,q)_{\mu\nu}\cM_{\rm 2PM}+\frac{i}{8}\int \hat{d}^{4}\ell\,\hat{\delta}\left(p_{1}\cdot\ell-\frac{\ell^{2}}{2}\right)\hat{\delta}\left(p_{2}\cdot\ell+\frac{\ell^{2}}{2}\right)\left[\mathcal{S}(\rho,q-\ell)_{\mu\nu}-\mathcal{S}(\rho,\ell)_{\mu\nu}\right]\cM_{t}^{\dagger}(q-\ell)\cM_{t}(\ell)\right\}.
\end{align}
\end{widetext}
The first term in curly brackets is the 2PM analog of \cref{eq:LOMemoryST}, and will be related to the portion of 2PM impulse transverse to the incoming momenta, $\mathcal{Q}_{1,{\rm 2PM}}^{\mu}$, while the second term is the cut contribution.
One must account for the full delta functions \cite{Kosower:2018adc,DiVecchia:2023frv,Caron-Huot:2023vxl}, and not only their linearized versions, because the cut contribution is superficially superclassical, so it must be expanded to subleading order in $\hbar$ to extract classical information.

Scrutinizing the cut contribution in more detail, we must be careful to correctly interpret $\cM^{\dagger}_{t}(q)$ when spin is involved.
Importantly, in contrast to the spinless case, $\cM^{\dagger}_{t}(q)\neq\left[\cM_{t}(q)\right]^{*}$.
Rather, for the $S$-matrix defined by $S=1+iT$,\footnote{The overall momentum-conserving delta function has the abbreviated argument $p=p_{1}+p_{2}-p_{1}^{\prime}-p_{2}^{\prime}$.}
\begin{align}
    \hat{\delta}^{D}(p)\cM^{\dagger}(q)&=\langle p_{1}^{\prime},p_{2}^{\prime}|T^{\dagger}|p_{1},p_{2}\rangle \\
    &=\left(\langle p_{1},p_{2}|T|p_{1}^{\prime},p_{2}^{\prime}\rangle\right)^{*}=\hat{\delta}^{D}(p)\left[\cM(-q)\right]^{*},\notag
\end{align}
where we can safely make the final identification in the classical limit, in which $\cO(\hbar)$ modifications of the massive momenta do not affect the scattering amplitude \cite{Haddad:2021znf,Aoude:2022trd}.
Specializing to tree-level two-to-two scattering, \cref{eq:1PMAmplitude}, we find
\begin{align}
    \cM^{\dagger}_{t}(q)=\cM_{t}(q).\label{eq:TdMatrixElement}
\end{align}
This conclusion is consistent with unitarity of the $S$-matrix, a consequence of which is that the $T$-matrix is Hermitian for the leading-order two-to-two scattering process.

Now, as the first quantum corrections to $\cM_{t}$ are suppressed by two powers of $\hbar$, the expansion of the cut contribution to next-to-leading order in $\hbar$ is controlled by the expansion of the product of the delta functions and soft factors.
A consequence of \cref{eq:TdMatrixElement} is that the super-classical part of this expansion vanishes at the level of the integrand, meaning \cref{eq:NLOMemoryST} is classical at leading order in $\hbar$.
Evaluating the remaining classical part of the cut contribution, the NLO memory effect expressed in terms of the 1PM and 2PM (transverse) impulses on particle 1 is
\begin{align}
    &\left.\Delta(h_{+}^{\infty}+ih_{\times}^{\infty})\right|^{\rm NLO}\label{eq:NLOMemoryResultImpulse}
    =\frac{\kappa\varepsilon^{\mu}_{-}\varepsilon^{\nu}_{-}}{8\pi|\boldsymbol{x}|}\left\{\mathcal{S}^{(0)}_{\mu\nu;\tau}(\rho)\mathcal{Q}^{\tau}_{1,{\rm 2PM}}\right. \\
    &-\mathcal{Q}_{1,{\rm 1PM}}^{\alpha}\mathcal{Q}_{1,{\rm 1PM}}^{\beta}\left[\mathcal{S}^{(0)}_{\mu\nu;\tau}(\rho)\left(\frac{\Check{v}_{1}^{\tau}}{2m_{1}}-\frac{\Check{v}_{2}^{\tau}}{2m_{2}}\right)\eta_{\alpha\beta}\right.\notag \\
    &\left.\left.+\mathcal{S}_{-\mu\nu;\alpha\beta}^{(1)}(\rho)\right]\right\},\notag
\end{align}
where
\begin{align}
    \Check{v}_{1}^{\mu}=\frac{\gamma v_{2}^{\mu}-v_{1}^{\mu}}{\gamma^{2}-1},\quad\Check{v}_{2}^{\mu}=\frac{\gamma v_{1}^{\mu}-v_{2}^{\mu}}{\gamma^{2}-1}.
\end{align}
This is in precise agreement with the gravitational memory (see, e.g., refs.~\cite{,Sahoo:2021ctw,DiVecchia:2022owy,DiVecchia:2022nna}) expanded to next-to-leading PM order, when the initial and final momenta are related by the classical impulse
\begin{align}
    Q^{\mu}&=p_{1}^{\prime\mu}-p_{1}^{\mu}=-(p_{2}^{\prime\mu}-p_{2}^{\mu}) \\
    &=\mathcal{Q}_{1,{\rm 1PM}}^{\mu}+\mathcal{Q}_{1,{\rm 2PM}}^{\mu}-\left(\frac{\Check{v}_{1}^{\mu}}{2m_{1}}-\frac{\Check{v}_{2}^{\mu}}{2m_{2}}\right)\mathcal{Q}_{1,{\rm 1PM}}^{2}.\notag
\end{align}
Without the cut contribution in \cref{eq:NLOMemoryST}, \cref{eq:NLOMemoryResultImpulse} would be missing the contributions quadratic in $\mathcal{Q}_{1,{\rm 1PM}}^{\mu}$. Note that, up the PM order considered here, we were able to focus on the so-called \emph{linear} memory, whose expression is captured by the soft factor \eqref{eq:LeadingSoftFactor} and in which only the massive-particle momenta $p_{1,2}$ and $p_{1,2}'$ appear. \emph{Non-linear} memory, which is produced by radiation itself, will only appear in the subsubleading waveform \cite{Damour:2020tta,DiVecchia:2022owy}.

As we have evaluated the 1PM impulse above, the last ingredient for writing the NLO memory effect explicitly is the evaluation of the 2PM transverse impulse.
We will do so for generic and anti-aligned spin orientations, beginning with the latter.

In the anti-aligned-spin setup, where $\mathfrak{a}_{\rm aa}\equiv\mathfrak{a}_{1}=-\mathfrak{a}_{2}$, the complexity of the all-spin amplitude is dramatically reduced, granting it a remarkably compact form to all spin orders.
The amplitude in this configuration is\footnote{\label{footnote:MappingComptons}The results in this section are based on the 2PM amplitude of ref.~\cite{Aoude:2023vdk} with $d_{j}^{(n)}=(-1)^{j}2^{n-2j}{{n-4-j}\choose{j}}-16\delta_{n4}$. This maps the Compton amplitude used there to construct the 2PM amplitude to the one incorporated in the waveform computation above, up to the contact terms in \cref{sec:ComptonDetails}. We add the latter separately.}
\begin{align}\label{eq:AmpAA}
    \cM_{2{\rm PM,aa}} &= \frac{\kappa^4 m^2_1 m^2_2}{512\sqrt{-q^2}} \left( \cM^{\rm even}_{2{\rm PM,aa}} + \cM^{\rm odd}_{2{\rm PM,aa}} \right),
\end{align}
where the even- and odd-in-spin parts, are
\begin{align}
    &\cM^{\rm even}_{2{\rm PM,aa}}= (m_{1}+m_{2})\left[\mathcal{C}_{2{\rm PM,aa}}^{{\rm even},(5)}+\mathcal{C}_{2{\rm PM,aa}}^{{\rm even},(6)}\right.\label{eq:NLOAAEven} \\
    &\left.+15(\gamma^{2}-1)\,{}_{2}F_{3}\left(-\frac{1}{4},\frac{1}{4};\frac{1}{2},\frac{3}{2},2;Q_{\rm aa}\right)-\frac{1}{2}Q_{\rm aa}+12\right]\notag, \\
    &\cM^{\rm odd}_{2{\rm PM,aa}}
    =i (m_1 - m_2) \gamma \left[\mathcal{C}_{2{\rm PM,aa}}^{{\rm odd},(5)}+\mathcal{C}_{2{\rm PM,aa}}^{{\rm odd},(6)}\right.\label{eq:NLOAAOdd} \\
    &\left. + 5\cE_{\rm aa}\;{}_{2}F_{3}\left(\frac{1}{4},\frac{3}{4};\frac{3}{2},2,\frac{5}{2};Q_{\rm aa}\right)+\frac{2}{\gamma^2 - 1}\cE_{\rm aa}\right]\notag.
\end{align}
We have abbreviated contributions from Compton-amplitude contact terms as $\mathcal{C}_{2{\rm PM,aa}}^{{\rm even/odd},(5,6)}$ for readability; see the ancillary Mathematica package \texttt{NLOMemory.m} for their explicit expressions.
When $v_{i}\cdot\mathfrak{a}_{j}=0$, which is the case for aligned and anti-aligned spins, all contributions to the 2PM amplitude from non-analytic-in-spin contact terms---that is, those proportional to $|\mathfrak{a}|$ in the Compton amplitude---vanish, as previously observed in refs.~\cite{Bautista:2022wjf,Bautista:2023szu}.

The spin dependence in \cref{eq:AmpAA} is encoded in the variables $Q_{\rm aa}=(q\cdot\mathfrak{a}_{\rm aa})^{2}-q^{2}\mathfrak{a}_{\rm aa}^{2}$ and $\cE_{\rm aa}=\epsilon_{\mu\nu\rho\sigma}q^{\mu}v_{1}^{\nu}v_{2}^{\rho}\mathfrak{a}_{\rm aa}^{\sigma}$.
The amplitude thus does not reduce to the Schwarzschild-scattering amplitude in this configuration, unlike the 1PM amplitude.
The NLO memory effect will therefore distinguish the scattering of two Schwarzschild black holes from two Kerr black holes with anti-aligned spins.
This statement is true independently of contact-term contributions, as contact terms enter from the hexadecapole while the memory effect is sensitive to lower spin multipoles.
Notably, however, the dependence on odd spin orders vanishes if we additionally take the two masses to be equal.

In this configuration the anti-aligned spins are collinear with the orbital angular momentum, which implies $b\cdot\mathfrak{a}_{\rm aa}=0$ and consequently allows us to find a compact closed form for the anti-aligned-spin transverse impulse at 2PM to all spin orders.
Defining 
$\cE_{\rm aa}^{\mu}\equiv \epsilon^{\mu\nu\alpha\beta}v_{1\nu}v_{2\alpha}\mathfrak{a}_{\rm aa,\beta}$, we find
\begin{widetext}
    \begin{align}
        &\mathcal{Q}^{\tau}_{1,\rm{2PM,aa}}=-\frac{\kappa^{4}m_{1}m_{2}}{4096\pi|b|^{3}\sqrt{\gamma^{2}-1}}\label{eq:NLOMemoryaa} \\
        &\times\left\{3b^{\tau}(m_{1}+m_{2})\left[4-z^{2}-\frac{5(\gamma^{2}-1)}{2\sqrt{\pi}}\sum_{n=0}^{\infty}\frac{\Gamma(2n-1/2)}{(n+1)}\frac{(2z)^{2n}}{(2n)!}+\tilde{\mathcal{C}}^{{\rm even},(5)}_{\rm NLO,aa}+\tilde{\mathcal{C}}^{{\rm even},(6)}_{\rm NLO,aa}\right]\right. \notag \\
        &+2(m_{1}-m_{2})\gamma\cE_{{\rm aa},\sigma}\left[\frac{1}{\gamma^{2}-1}\left(3\frac{b^{\sigma}b^{\tau}}{|b|^{2}}+\eta^{\sigma\tau}\right)+\frac{15}{\sqrt{\pi}}\sum_{n=0}^{\infty}\frac{\Gamma(2n+1/2)}{(2n+3)!}\left((2n+3)\frac{b^{\sigma}b^{\tau}}{|b|^{2}}+\eta^{\sigma\tau}\right)(2z)^{2n}\right.\notag \\
        &\left.\left.+\left(\tilde{\mathcal{C}}^{{\rm odd},(5)}_{\rm NLO,aa}\right)^{\sigma\tau}+\left(\tilde{\mathcal{C}}^{{\rm odd},(6)}_{\rm NLO,aa}\right)^{\sigma\tau}\right]\right\},\notag
    \end{align}
\end{widetext}
where $z\equiv |\mathfrak{a}_{\rm aa}|/|b|$, $|x|\equiv\sqrt{-x^{2}}$.
The infinite sums can be performed to give radicals and hypergeometric functions, but we find this expression to be more compact.
The portions involving Compton-amplitude contact coefficients can be found in the ancillary file \texttt{NLOMemory.m}.

The anti-aligned-spin configuration is an interesting one as the simplifications it brings with it render the infinite-spin 2PM amplitude much more manageable.
Phenomenologically, however, it is a rather restrictive setup.
For this reason, we additionally consider the NLO memory effect for generically oriented spins.
We restrict our attention up to sixth order in spin in this most general configuration.
Analogously to \cref{eq:AmpAA}, we write the amplitude up to sixth order in spin as
\begin{align}
    \left.\cM_{2{\rm PM}}\right|_{\mathfrak{a}^{n\leq6}} &= \frac{\kappa^4 m^2_1 m^2_2}{512\sqrt{-q^2}} \left.\left( \cM^{\rm even}_{2{\rm PM}} + \cM^{\rm odd}_{2{\rm PM}} \right)\right|_{\mathfrak{a}^{n\leq6}},\label{eq:Amplitude2PMGenericOrientations}
\end{align}
while the 2PM transverse impulse on particle 1 becomes
\begin{align}
    &\left.\mathcal{Q}_{1,{\rm 2PM}}^{\tau}\right|_{\mathfrak{a}^{n\leq6}}=-\frac{\kappa^{4}m_{1}m_{2}\left.\left(\mathfrak{q}_{1,{\rm 2PM}}^{{\rm even,}\tau}+\mathfrak{q}_{1,{\rm 2PM}}^{{\rm odd,}\tau}\right)\right|_{\mathfrak{a}^{n\leq6}}}{4096\pi|b|^{3}\sqrt{\gamma^{2}-1}}.\label{eq:TransverseImpulse2PMGenericOrientations}
\end{align}
In this configuration we relegate all further analytical details of the amplitude and the impulse to the ancillary file \texttt{NLOMemory.m}.
The amplitude in \cref{eq:Amplitude2PMGenericOrientations} and transverse 2PM impulse in \cref{eq:TransverseImpulse2PMGenericOrientations} are in agreement with ref.~\cite{Bautista:2023szu} for the BHPT coefficient values in \cref{sec:ComptonDetails}.

\section{Conclusion}
\label{sec:Conclusions}

In this paper, we have employed on-shell amplitudes and spinor-helicity variables to access all-spin-order contributions to the leading-order waveform and the gravitational memory effect up to next-to-leading order.
In particular, gluing the Kerr-compatible, all-spin gravitational Compton amplitude derived in ref.~\cite{Aoude:2022trd} (but in the form written in ref.~\cite{Aoude:2022thd}) with the all-spin Kerr three-point amplitude \cite{Levi:2015msa,Arkani-Hamed:2017jhn,Vines:2017hyw,Guevara:2018wpp,Chung:2018kqs} gives the portion of the single-graviton-emission, five-point amplitude containing long-distance information to all spin orders.

The KMOC formalism \cite{Kosower:2018adc,Cristofoli:2021vyo} provides a means for relating this portion of the five-point amplitude to the leading-order gravitational waveform, producing an expression for the waveform, which is valid to all spin orders.
As written, \cref{eq:WaveformImplicit} describes the interaction of Kerr black holes up to fourth order in the spins of either black hole.
Above fourth order in spin, contributions from Compton-amplitude contact terms are needed to properly describe Kerr scattering dynamics.
Indeed, our analysis included these corrections up to sixth order in spin in \cref{eq:WaveformContacts}, where information from BHPT exists to fix the coefficient values pertinent to (super-extremal) Kerr \cite{Bautista:2022wjf}.
Together, \cref{eq:WaveformImplicit,eq:WaveformContacts} sum as in \cref{eq:WaveformDecomposed} to describe the leading-order Kerr waveform---at least in the super-extremal limit---up to sixth order in the spins of the black holes.
We have written the cut portion of the five-point amplitude in \cref{eq:FivePointAmplitude} in a way that immediately accommodates higher-spin contact terms, and \cref{eq:WaveformContacts} is not difficult to extend to include such contributions.

The gravitational memory effect can be extracted from the limit of the five-point amplitude needed for the waveform as the emitted graviton goes soft.
Then, through soft theorems \cite{Weinberg:1965nx}, including also the cut contribution to the waveform kernel of ref.~\cite{Caron-Huot:2023vxl}, it becomes easily related to the impulse derived from the amplitude through the KMOC formalism \cite{Kosower:2018adc}.
Using the all-spin 1PM Kerr \cite{Guevara:2019fsj} and 2PM Kerr-compatible \cite{Aoude:2023vdk} amplitudes, we thus derived the leading-order memory effect to all spin orders at leading order and to sixth order in spin at next-to-leading order for generic spin orientations.
Specializing to anti-aligned spins yielded dramatic simplifications of the 2PM amplitude, enabling us to extract the next-to-leading-order memory effect to all spin orders in this configuration.

On the note of contact terms, those needed in \cref{eq:ContactFFOH0,eq:ContactFFOH1,eq:ContactFFOH2,eq:ContactFFOH3,eq:ContactFFOH4} to match the BHPT solution in ref.~\cite{Bautista:2022wjf} all break the spin-shift symmetry highlighted in refs.~\cite{Aoude:2022trd,Bern:2022kto,Aoude:2022thd}.
This observation is suggestive of a relation between the Compton amplitude in the form of ref.~\cite{Aoude:2022thd} with all contact coefficients set to zero and the super-extremal-Kerr Compton amplitude which maps onto the BHPT description.
Let us denote the former by $\cM_{4}^{\rm HPET}$ and decompose the latter as
\begin{align}
    \cM_{4}^{\rm BHPT}=\cM_{4}^{\rm fact.}+\mathcal{C}_{4}^{\rm sym.}+\mathcal{C}_{4}^{\rm asym.}.
\end{align}
Here, $\cM_{4}^{\rm fact.}$ contains all physical residues of the Compton amplitude, $\mathcal{C}_{4}^{\rm sym.}$ represents contact terms preserving the spin-shift symmetry, and $\mathcal{C}_{4}^{\rm asym.}$ contains contact terms breaking this symmetry.\footnote{In principle, the spin-shift-symmetric and asymmetric contact terms can be mixed by Gram determinant relations as in ref.~\cite{Aoude:2022trd}. However, we observe that this separation is well-defined for the contact terms appearing in the BHPT solution if one writes the amplitude manifestly locally.}
We reiterate that the latter two are not generic functions of contact terms, but rather the specific contact terms arising from the BHPT computation.
The separation between $\cM_{4}^{\rm fact.}$ and $\mathcal{C}_{4}^{\rm sym.}$ is not unique, but their sum is fixed.
What we have observed up to $\cO(\mathfrak{a}^{7})$\footnote{The comparison between $\cM_{4}^{\rm BHPT}$ and $\cM_{4}^{\rm HPET}$ at $\cO(\mathfrak{a}^{7})$ was done using unpublished data generously shared by Fabian Bautista.} and might conjecture to hold to all spin orders is that
\begin{align}\label{eq:ShiftSymmetricKerrConjecture}
    \cM_{4}^{\rm fact.}+\mathcal{C}_{4}^{\rm sym.}=\cM_{4}^{\rm HPET},
\end{align}
thus yielding predictions for the values of an infinite family of the contact terms needed to match the BHPT description of super-extremal Kerr.
Whether this regrouping of contact terms results in a discernible structure in $\mathcal{C}_{4}^{\rm asym.}$ which can be extended to higher spins is left to future investigation.

Along similar lines, the extremely compact form of \cref{eq:AmpAA} suggests that the anti-aligned spin configuration may be a useful departure point in the search for contact-term-dependent structure of the amplitude proposed in ref.~\cite{Aoude:2023vdk}.
In fact, we have observed that choosing the shift-symmetric contact terms conjectured by \cref{eq:ShiftSymmetricKerrConjecture} to describe Kerr black holes---that is, the contact terms specified in \cref{footnote:MappingComptons}---compactifies \cref{eq:NLOAAEven} relative to the choice $d_{j}^{(n)}=0$; in the latter case, the spin dependence is described by two hypergeometric functions as opposed to one.

\subsection*{Acknowledgments}

This project was inspired by Donal O'Connell's talk at the Nordita workshop ``Amplifying Gravity at All Scales."
We thank Fabian Bautista, Zvi Bern, Stefano De Angelis, Gustav Uhre Jakobsen, Gustav Mogull, Jan Plefka, Rodolfo Russo, and Biswajit Sahoo for helpful discussions.
We are grateful to Stefano De Angelis, Riccardo Gonzo, and Pavel Novichkov for their open communication during their work on ref.~\cite{DeAngelis:2023lvf} and for sharing a draft before its release.
R.A. is supported by the FSR Program of UCLouvain.
K. H. is supported by the Knut and Alice Wallenberg Foundation under grants KAW 2018.0116 (From Scattering Amplitudes to Gravitational Waves) and KAW 2018.0162.
K. H. is grateful to Nordita for their ongoing hospitality.
C. H. is supported by UK Research and Innovation (UKRI) under the UK government’s Horizon Europe funding guarantee [grant EP/X037312/1 ``EikoGrav: Eikonal Exponentiation and Gravitational Waves''].
A. H. is supported by the DOE under award number DE-SC0011632 and by the Walter Burke Institute for Theoretical Physics.

\appendix

\section{Form factors and helicity vectors}\label{sec:ComptonDetails}

The tensors $r_{(i),n}^{h,\mu\nu}$ which carry the helicity weights of the five-point amplitude in \cref{eq:FivePointAmplitude} are
\begin{subequations}\label{eq:HelicityVectors}
\begin{align}
    r_{(1),0}^{+,\mu\nu}&=\langle k|p_{1}p_{2}\gamma^{\mu}p_{1}|k\rangle\langle k|p_{1}p_{2}\gamma^{\nu}p_{1}|k\rangle, \\
    r_{(1),1}^{+,\mu\nu}&=\langle k|p_{1}p_{2}\gamma^{\mu}p_{1}|k\rangle\langle k|p_{1}p_{2}\gamma^{\nu}\mathfrak{a}_{1}|k\rangle, \\
    r_{(1),2}^{+,\mu\nu}&=\langle k|p_{1}p_{2}\gamma^{\mu}\mathfrak{a}_{1}|k\rangle\langle k|p_{1}p_{2}\gamma^{\nu}\mathfrak{a}_{1}|k\rangle, \\
    r_{(1),3}^{+,\mu\nu}&=\langle k|p_{1}p_{2}\gamma^{\mu}\mathfrak{a}_{1}|k\rangle\langle k|\mathfrak{a}_{1}p_{2}\gamma^{\nu}\mathfrak{a}_{1}|k\rangle, \\
    r_{(1),4}^{+,\mu\nu}&=\langle k|\mathfrak{a}_{1}p_{2}\gamma^{\mu}\mathfrak{a}_{1}|k\rangle\langle k|\mathfrak{a}_{1}p_{2}\gamma^{\nu}\mathfrak{a}_{1}|k\rangle,
\end{align}
and
\begin{align}
    r_{(1),0}^{-,\mu\nu}&=m_{1}^{4}\langle k|p_{2}\gamma^{\mu}|k\rangle\langle k|p_{2}\gamma^{\nu}|k\rangle, \\
    r_{(1),1}^{-,\mu\nu}&=m_{1}^{2}\langle k|p_{2}\gamma^{\mu}|k\rangle\langle k|\mathfrak{a}_{1}p_{1}p_{2}\gamma^{\nu}|k\rangle, \\
    r_{(1),2}^{-,\mu\nu}&=\langle k|\mathfrak{a}_{1}p_{1}p_{2}\gamma^{\mu}|k\rangle\langle k|\mathfrak{a}_{1}p_{1}p_{2}\gamma^{\nu}|k\rangle, \\
    r_{(1),3}^{-,\mu\nu}&=\frac{1}{m_{1}^{2}}\langle k|\mathfrak{a}_{1}p_{1}p_{2}\gamma^{\mu}|k\rangle\langle k|\mathfrak{a}_{1}p_{1}p_{2}\gamma^{\nu}p_{1}\mathfrak{a}_{1}|k\rangle, \\
    r_{(1),4}^{-,\mu\nu}&=\frac{1}{m_{1}^{4}}\langle k|\mathfrak{a}_{1}p_{1}p_{2}\gamma^{\mu}p_{1}\mathfrak{a}_{1}|k\rangle\langle k|\mathfrak{a}_{1}p_{1}p_{2}\gamma^{\nu}p_{1}\mathfrak{a}_{1}|k\rangle. 
\end{align}
\end{subequations}
The $r_{(2),n}^{h,\mu\nu}$ are obtained from these using \cref{eq:RelabellingNotation}.
When a positive-helicity graviton is emitted from the binary scattering, the amplitude will depend on $\bar{r}_{(i),n}^{-h,\mu\nu}$ in place of $r_{(i),n}^{h,\mu\nu}$, which have square spinors rather than the angle spinors above.
These tensors are already inert under the waveform integration over the $q_{i}$; we can render them inert under the integration over $\omega$ as well by noting that $r_{(i),n}^{h,\mu\nu}=\omega^{2}\tilde{r}_{(i),n}^{h,\mu\nu}$, where the tensors on the right-hand side are written with spinors for $\rho$ instead of $k$.
The same holds for the tensors with square instead of angle brackets.

The $F^{(n)}_{4}$ form factors are
\begin{align}
    F_{4}^{(0)}(-p,k^{h},-q^{h})&=\frac{1}{8(q\cdot k)(p\cdot k)(p\cdot q)},\label{eq:FactorizableFFSH1} \\
    F_{4}^{(n\geq1)}(-p,k^{h},-q^{h})&=0,\label{eq:FactorizableFFSH2}
\end{align}
and, fixing the helicity of the graviton with momentum $q$,
\begin{align}
    &F_{4}^{(n\leq2)}(-p,k^{-},-q^{+})=\frac{[2p\cdot(q+k)]^{n}}{8n!(q\cdot k)(p\cdot k)(p\cdot q)},\label{eq:FactorizableFFOH1} \\
    &F_{4}^{(n\geq3)}(-p,k^{-},-q^{+})\label{eq:FactorizableFFOH2} \\
    &=\frac{[2p\cdot(q+k)]^{n}}{8(q\cdot k)(p\cdot k)(p\cdot q)}\left[\frac{1}{n!}+(-\mathfrak{s}_{2})^{4-n}\sum_{s=5}^{\infty}\frac{1}{s!}L_{s-8+n}\right]\notag,
\end{align}
where
\begin{align}
	L_{m} =& \sum_{j=0}^{\floor{m/2}} \binom{m+1}{2j+1} \mathfrak{s}_{1}^{m-2j} \left( \mathfrak{s}_{1}^{2} - \mathfrak{s}_{2} \right)^{j} \,,
\end{align}
and 
\begin{align}
	\mathfrak{s}_{1} &= (k + q) \cdot \mathfrak{a} \,, \\
	\mathfrak{s}_{2} &= 4 (k \cdot \mathfrak{a})(q \cdot \mathfrak{a}) - (2q\cdot k) \mathfrak{a}^2 \,.
\end{align}
The form factors for $F_{4}^{(n)}(-p,k^{+},-q^{-})$ are obtained from \cref{eq:FactorizableFFOH1,eq:FactorizableFFOH2} by replacing $\{k^{\mu},q^{\mu}\}\rightarrow\{-k^{\mu},-q^{\mu}\}$.

As the extraction of the waveform involves integrals over momenta represented here by $q^{\mu}$, it is useful to rewrite $F_{4}^{(n\geq3)}(-p,k^{-},-q^{+})$ as an expansion in $q^{\mu}$, which entails expanding $L_{m}$ as such.
This actually becomes easier after integrating over one of the $\hat{d}^{4}q_{i}$ and $\hat{d}\omega$ using a delta function as described in \cref{sec:Waveform}.
For example, let us consider the $L_{m}$ contributing to the $q_{2}^{2}$ pole, which we write as $L_{(1),m}$.
After integrating over $\hat{d}^{4}q_{1}$ and $\hat{d}\omega$, this becomes
\begin{align}
    \left.L_{(1),m}\right|_{\omega=\frac{v_{1}\cdot q_{2}}{v_{1}\cdot\rho}}=q_{2\mu_{1}}\dots q_{2\mu_{m}}L_{(1),m}^{\mu_{1}\dots\mu_{m}},
\end{align}
where
\begin{widetext}
    \begin{align}
        L_{(1),m}^{\mu_{1}\dots\mu_{m}}&\equiv\sum_{j=0}^{\floor{m/2}} \binom{m+1}{2j+1}\prod_{i=1}^{m-2j}\left[\frac{\rho\cdot\mathfrak{a}_{1}}{v_{1}\cdot\rho}v_{1}^{\mu_{i}}+\mathfrak{a}_{1}^{\mu_{i}}\right] \\
        &\quad\times\prod_{k=\frac{m-2j+1}{2}}^{(m-1)/2}\left[\left(\frac{\rho\cdot\mathfrak{a}_{1}}{v_{1}\cdot\rho}v_{1}^{\mu_{2k}}-\mathfrak{a}_{1}^{\mu_{2k}}\right)\left(\frac{\rho\cdot\mathfrak{a}_{1}}{v_{1}\cdot\rho}v_{1}^{\mu_{2k+1}}-\mathfrak{a}_{1}^{\mu_{2k+1}}\right) + 2v_{1}^{\mu_{2k}}\rho^{\mu_{2k+1}}\frac{\mathfrak{a}^{2}_{1}}{v_{1}\cdot\rho}\right],\notag
    \end{align}
    which subsequently enters the waveform through
    \begin{align}\label{eq:FormFactorNoq}
        \mathcal{L}_{(1),s}^{\mu_{1}\dots\mu_{s+2}}&=\begin{cases}
            m_{1}^{s}\tilde{r}_{(1),s}^{+,\mu_{1}\mu_{2}}(2v_{1}^{\mu_{3}})\dots (2v_{1}^{\mu_{s+2}}), & s\leq 4, \\
            (2v_{1}^{\mu_{3}})\dots (2v_{1}^{\mu_{6}})\left[m_{1}^{4}\tilde{r}_{(1),4}^{+,\mu_{1}\mu_{2}}L_{(1),s-4}^{\mu_{7}\dots\mu_{s+2}}-m_{1}^{3}\tilde{r}_{(1),3}^{+,\mu_{1}\mu_{2}}\left(2\frac{\rho \cdot \mathfrak{a}_{1}}{v_{1}\cdot\rho}\mathfrak{a}_{1}^{\mu_{7}} - \rho^{\mu_{7}}\frac{\mathfrak{a}^{2}_{1}}{v_{1}\cdot\rho}\right)L_{(1),s-5}^{\mu_{8}\dots\mu_{s+2}}\right], & s>4.
        \end{cases}
    \end{align}
\end{widetext}
The tensors $L_{(2),m}^{\mu_{1}\dots\mu_{m}}$ and $\mathcal{L}_{(2),s}^{\mu_{1}\dots\mu_{s+2}}$ can be obtained from these by swapping the labels $1\leftrightarrow2$.
Note that $L_{(i),0}=1$.

Instead of computing with the full set of contact terms compatible with Kerr scattering at the PM order considered, we will focus only on those entering up to sixth order in spin.
Moreover we will fix to zero all contact term coefficients that are not needed to match the super-extremal analytic continuation of the BHPT solution in ref.~\cite{Bautista:2022wjf}.
This means we consider
\begin{widetext}
\begin{align}
    &C_{4}^{(n)}(-p,k^{-},-q^{-})=0,\label{eq:ContactFFSH} \\
    &C_{4}^{(0)}(-p,k^{-},-q^{+})=\frac{(t_{pq}-t_{pk})}{m^{3}}|\mathfrak{a}|\mathfrak{a}^{4}\left[f_{0,0,0}^{(5)}+f_{0,0,1}^{(6)}[(-q-k)\cdot\mathfrak{a}]\right]\label{eq:ContactFFOH0} \\
    &\quad+a_{1,0,0}^{(6)}\,\frac{s_{qk}}{m^{2}}\mathfrak{a}^{6}+a_{1,1,0}^{(6)}\,\frac{(t_{pq}-t_{pk})^{2}}{m^{4}}\mathfrak{a}^{6}-2\left(a_{0,0,0}^{(6)}-a_{0,0,2}^{(6)}\right)(q\cdot\mathfrak{a})(k\cdot\mathfrak{a})\frac{\mathfrak{a}^{4}}{m^{2}},\notag \\
    &C_{4}^{(1)}(-p,k^{-},-q^{+})=\frac{(t_{pq}-t_{pk})}{m^{2}}\mathfrak{a}^{4}\left[b_{0,0,0}^{(5)}+b_{0,0,1}^{(6)}[(-q-k)\cdot\mathfrak{a}]\right]\label{eq:ContactFFOH1} \\
    &\quad+g_{1,0,0}^{(6)}\frac{s_{qk}}{m}|\mathfrak{a}|\mathfrak{a}^{4}+g_{1,1,0}^{(6)}\,\frac{(t_{pq}-t_{pk})^{2}}{m^{3}}|\mathfrak{a}|\mathfrak{a}^{4}-2\left(g_{0,0,0}^{(6)}-g_{0,0,2}^{(6)}\right)\frac{1}{m}|\mathfrak{a}|\mathfrak{a}^{2}(q\cdot\mathfrak{a})(k\cdot\mathfrak{a}),\notag \\
    &C_{4}^{(2)}(-p,k^{-},-q^{+})= \frac{(t_{pq}-t_{pk})}{m}|\mathfrak{a}|\mathfrak{a}^{2}\left[p_{0,0}^{(5)}+p_{0,1}^{(6)}[(-q-k)\cdot\mathfrak{a}]\right] + c_{1,0}^{(6)}\,s_{kq}\mathfrak{a}^{4}-2\left(c_{0,0}^{(6)}-c_{0,2}^{(6)}\right)(q\cdot\mathfrak{a})(k\cdot\mathfrak{a})\mathfrak{a}^{2},\label{eq:ContactFFOH2} \\
    &C_{4}^{(3)}(-p,k^{-},-q^{+})=(t_{pq}-t_{pk})\mathfrak{a}^{2}\left[d_{0,0}^{(5)}+d_{0,1}^{(6)}[(-q-k)\cdot\mathfrak{a}]\right]+q_{1,0}^{(6)}\,ms_{qk}|\mathfrak{a}|\mathfrak{a}^{2},\label{eq:ContactFFOH3} \\
    &C_{4}^{(4)}(-p,k^{-},-q^{+})=m(t_{pq}-t_{pk})|\mathfrak{a}|\left[r_{0,0}^{(5)}+r_{0,1}^{(6)}[(-q-k)\cdot\mathfrak{a}]\right] + e_{1,0}^{(6)}\,m^{2}s_{kq}\mathfrak{a}^{2}.\label{eq:ContactFFOH4}
\end{align}
\end{widetext}
We have defined $s_{qk}\equiv(k-q)^{2}$, $t_{pq}\equiv(p+q)^{2}-m^{2}$, and $t_{pk}\equiv(p-k)^{2}-m^{2}$.
The coefficients used here are those from ref.~\cite{Aoude:2023vdk}, and their values matching the super-extremal solution to the Teukolsky equation at fifth order in spin, according to ref.~\cite{Bautista:2022wjf}, are
\begin{align}
    b_{0,0,0}^{(5)}=-\tfrac{1}{24},&\quad d_{0,0}^{(5)}=\tfrac{1}{3},\notag \\
    f_{0,0,0}^{(5)}=-\tfrac{1}{240},&\quad p_{0,0}^{(5)}=\tfrac{1}{6}, \\
    r_{0,0}^{(5)}=-\tfrac{1}{3}.\notag
\end{align}
At sixth order in spin, the coefficient values in \cref{eq:FivePointAmplitude} matching ref.~\cite{Bautista:2022wjf} are
\begin{align}
    a_{1,0,0}^{(6)}=-\tfrac{1}{16},&\quad a_{1,1,0}^{(6)}=-\tfrac{1}{576},\notag \\
    a_{0,0,0}^{(6)}-a_{0,0,2}^{(6)}=\tfrac{1}{8},&\quad b_{0,0,1}^{(6)}=-\tfrac{11}{72},\notag \\
    c_{0,0}^{(6)}-c_{0,2}^{(6)}=-\tfrac{1}{6},&\quad c_{1,0}^{(6)}=-\tfrac{1}{6},\notag \\
    d_{0,1}^{(6)}=\tfrac{7}{18},&\quad e_{1,0}^{(6)}=\tfrac{1}{3}, \\
    f_{0,0,1}^{(6)}=-\tfrac{1}{360},&\quad g_{1,0,0}^{(6)}=\tfrac{1}{9},\notag \\
    g_{0,0,0}^{(6)}-g_{0,0,2}^{(6)}=-\tfrac{2}{9},&\quad g_{1,1,0}^{(6)}=\tfrac{1}{60},\notag \\
    p_{0,1}^{(6)}=\tfrac{1}{3},\quad q_{1,0}^{(6)}&=\tfrac{4}{9},\quad r_{0,1}^{(6)}=-\tfrac{2}{9}.\notag
\end{align}
It is immediate to augment \cref{eq:FivePointAmplitude} with more Compton-amplitude contact terms since the $C_{4}^{(n)}$ are inert under the gluing of the three-point and Compton amplitudes in \cref{fig:FivePointCuts}.

The contact terms repackaged in preparation for the waveform integration are encoded in the form factors $C_{4}^{(j,k)}(\mathfrak{a}_{i})$ introduced in \cref{eq:WaveformContacts}.
Written explicitly, these are
\begin{widetext}
    \begin{align}
        C_{4}^{(5,1),\mu\nu}(\mathfrak{a}_{i})&=\frac{\tilde{r}^{+,\mu\nu}_{(i),0}}{m^{3}_{i}}f_{0,0,0}^{(5)}|\mathfrak{a}_{i}|\mathfrak{a}^{4}_{i}+\frac{\tilde{r}^{+,\mu\nu}_{(i),1}}{2m^{2}_{i}}b_{0,0,0}^{(5)}\mathfrak{a}^{4}_{i}+\frac{\tilde{r}^{+,\mu\nu}_{(i),2}}{4m_{i}}p_{0,0}^{(5)}|\mathfrak{a}_{i}|\mathfrak{a}_{i}^{2}+\frac{\tilde{r}^{+,\mu\nu}_{(i),3}}{8}d_{0,0}^{(5)}\mathfrak{a}_{i}^{2}+m_{i}\frac{\tilde{r}^{+,\mu\nu}_{(i),4}}{16}r_{0,0}^{(5)}|\mathfrak{a}_{i}|,\label{eq:RepackagedContact1} \\
        C_{4}^{(6,1),\mu\nu\alpha}(\mathfrak{a}_{i})&=4p_{i}^{\alpha}\left(\frac{\tilde{r}^{+,\mu\nu}_{(i),0}}{m^{4}_{i}}a_{1,1,0}^{(6)}\mathfrak{a}_{i}^{6}+\frac{\tilde{r}^{+,\mu\nu}_{(i),1}}{2m_{i}^{3}}g_{1,1,0}^{(6)}|\mathfrak{a}_{i}|\mathfrak{a}_{i}^{4}\right),\label{eq:RepackagedContact2} \\
        C_{4}^{(6,2),\mu\nu\alpha}(\mathfrak{a}_{i})&=-\left(\mathfrak{a}_{i}^{\alpha}+\frac{p_{i}^{\alpha}\rho\cdot\mathfrak{a}_{i}}{p_{i}\cdot\rho}\right)\label{eq:RepackagedContact3}  \\
        &\quad\times\left(\frac{\tilde{r}^{+,\mu\nu}_{(i),0}}{m_{i}^{3}}f_{0,0,1}^{(6)}|\mathfrak{a}_{i}|\mathfrak{a}_{i}^{4}+\frac{\tilde{r}^{+,\mu\nu}_{(i),1}}{2m_{i}^{2}}b_{0,0,1}^{(6)}\mathfrak{a}_{i}^{4}+\frac{\tilde{r}^{+,\mu\nu}_{(i),2}}{4m_{i}}p_{0,1}^{(6)}|\mathfrak{a}_{i}|\mathfrak{a}_{i}^{2}+\frac{\tilde{r}^{+,\mu\nu}_{(i),3}}{8}d_{0,1}^{(6)}\mathfrak{a}_{i}^{2}+m_{i}\frac{\tilde{r}^{+,\mu\nu}_{(i),4}}{16}r_{0,1}^{(6)}|\mathfrak{a}_{i}|\right),\notag\\
        C_{4}^{(6,3),\mu\nu\alpha}(\mathfrak{a}_{i})&=\frac{\rho_{\beta}}{4p_{1}\cdot\rho}\left[-2\frac{\tilde{r}^{+,\mu\nu}_{(i),0}}{m_{i}^{2}}\mathfrak{a}_{i}^{4}\left(a_{0,0,0}^{(6)}-a_{0,0,2}^{(6)}\right)\mathfrak{a}^{\alpha}_{i}\mathfrak{a}_{i}^{\beta}-2\frac{\tilde{r}^{+,\mu\nu}_{(i),1}}{2m_{i}}|\mathfrak{a}_{i}|\mathfrak{a}_{i}^{2}\left(g_{0,0,0}^{(6)}-g_{0,0,2}^{(6)}\right)\mathfrak{a}^{\alpha}_{i}\mathfrak{a}_{i}^{\beta}\right.\label{eq:RepackagedContact4} \\
        &\left.-2\frac{\tilde{r}^{+,\mu\nu}_{(i),2}}{4}\mathfrak{a}_{i}^{2}\left[c_{0,1}^{(6)}\eta^{\alpha\beta}\mathfrak{a}_{i}^{2}+\left(c_{0,0}^{(6)}-c_{0,2}^{(6)}\right)\mathfrak{a}^{\alpha}_{i}\mathfrak{a}_{i}^{\beta}\right]-2m_{i}\frac{\tilde{r}^{+,\mu\nu}_{(i),3}}{8}q_{1,0}^{(6)}\eta^{\alpha\beta}|\mathfrak{a}_{i}|\mathfrak{a}_{i}^{2}-2m_{i}^{2}\frac{\tilde{r}^{+,\mu\nu}_{(i),4}}{16}e_{1,0}^{(6)}\eta^{\alpha\beta}\mathfrak{a}_{i}^{2}\right].\notag
    \end{align}
\end{widetext}

\section{Integrals}\label{sec:Integrals}

There are two classes of integrals that we must compute to convert from momentum to impact-parameter space: one for the evaluation of the waveform and one for the memory effect.
These are
\begin{align}
    &I^{\mu_{1}\dots\mu_{n}}_{w}[\{x,b_{1},b_{2}\};f(q)]=\int \hat d\omega \hat{d}^{4}q\hat{d}^{4}q^{\prime}\, \hat{\delta}(v_{1}\cdot q)\hat{\delta}(v_{2}\cdot q^{\prime})\notag \\
    &\times\hat{\delta}^{(4)}(k-q-q^{\prime})e^{-i\omega\rho\cdot x}e^{i(q\cdot b_1 + q^\prime \cdot b_2)}q^{\mu_{1}}\dots q^{\mu_{n}}f(q), \\
    &I^{\mu_{1}\dots\mu_{n}}_{m}[b;f(q)]=\int \hat{d}^{4}q\, \hat{\delta}(v_{1}\cdot q)\hat{\delta}(v_{2}\cdot q)\notag \\
    &\times e^{iq\cdot b}q^{\mu_{1}}\dots q^{\mu_{n}}f(q),
\end{align}
respectively.
For amplitudes involving arbitrary spin powers, these integrals generally must be evaluated for arbitrary rank.
Instead of evaluating each rank individually, higher-rank integrals can be generated from lower ranks by differentiation.

\subsection{Waveform integration}

Extracting the waveform in an explicit form requires that we evaluate the integrals in \cref{eq:WaveformFactorizableIntegral}:
\begin{align}
	&\frac{\cI^{\mu_1 \cdots \mu_n}_{(1)}(b_{(1),\pm})}{v_{1}\cdot\rho}= \\
    &I_{w}^{\mu_1 \cdots \mu_n}\left[\{x-i\mathfrak{a}_{1},b_{1},b_{2}\pm i(\mathfrak{a}_{1}+\mathfrak{a}_{2})\};\frac{1}{q^2_{2} (q_{2}\cdot \rho)(v_1 \cdot q_{2})} \right] .\notag
\end{align}
The lowest-rank integral we need is \cite{Jakobsen:2021lvp}
\begin{align}
	\label{eq:wfInt}
    \cI^{\mu\nu}_{(1)}(b) &=\frac{ K_{(1)}^{\mu\nu} (v_1 \cdot K_{(1)} \cdot \rho) - 2 (v_1 \cdot K_{(1)})^{(\mu} (\rho \cdot K_{(1)} )^{\nu)} }{4\pi(\gamma^2 - 1)(\rho \cdot v_2)^2 |b|^{2} |\boldsymbol{b}|_{(1)}|b|_{\rm 2d}^{2}}  ,
\end{align}
where
\begin{align}
    K_{(1)}^{\mu\nu} &= \Pi_{{\rm 3d},(1)}^{\mu\nu} |\boldsymbol{b}|_{(1)}^2 + \boldsymbol{b}_{(1)}^{\mu} \boldsymbol{b}_{(1)}^{\nu} ,\notag \\
	\Pi_{{\rm 3d},(1)}^{\mu\nu} &= \eta^{\mu\nu} - v_2^{\mu} v_2^{\nu},\quad \Pi^{\mu\nu}_{{\rm 2d}} = \eta^{\mu\nu}-v_{1}^{\mu}\Check{v}_{1}^{\nu}-v_{2}^{\mu}\Check{v}_{2}^{\nu},  \nonumber \\
    \boldsymbol{b}^{\mu}_{(1)}&=\Pi^{\mu\nu}_{{\rm 3d},(1)}b_{\nu},\quad |\boldsymbol{b}|_{(1)}^2 = - b_{\mu}b_{\nu}\Pi_{{\rm 3d},(1)}^{\mu\nu}, \\
    |b|_{\rm 2d}^2 &= - b_{\mu}b_{\nu}\Pi^{\mu\nu}_{\rm 2d},\quad |b|^2 = - b_\mu b^{\mu}.\notag
\end{align}
An important feature of the integral in \cref{eq:wfInt} is that it is traceless, $\eta_{\mu\nu} \cI^{\mu\nu}_{(i)} = 0$.
This means that contributions from parts of the amplitude with spurious poles but no physical graviton poles don't contribute to the waveform.
Said otherwise, this justifies our use of the cut amplitude in \cref{sec:FivePointAmplitude} (which has unphysical poles in $q_{i}\cdot k$) for the extraction of the waveform rather than the whole amplitude.

All higher-rank integrals can be obtained from \cref{eq:wfInt} by differentiating with respect to $b$, with the constraint that the result should remain orthogonal to $v_2$; that is to say that we differentiate with respect to $\boldsymbol{b}^{\mu}_{(1)}$.
For the most part, this is straightforward, using
\begin{equation}
\begin{aligned}
    \frac{\partial\boldsymbol{b}_{(1)}^{\mu}}{\partial \boldsymbol{b}_{(1),\rho}}&=\Pi^{\mu\rho}_{{\rm 3d},(1)}, \\
    \frac{\partial}{\partial \boldsymbol{b}_{(1),\rho}}\left(\Pi^{\mu\nu}_{\rm 2d}b_{\nu}\right)&=\Pi^{\mu\rho}_{\rm 2d}.
\end{aligned}
\end{equation}
The derivative of $|b|^{2}$ is more involved. 
We must write $b^{\mu}$ in terms of $\boldsymbol{b}^{\mu}_{(1)}$, considering that $\rho\cdot b=v_{2}\cdot\boldsymbol{b}_{(1)}=0$:
\begin{align}
    b^{\mu}=\boldsymbol{b}^{\mu}_{(1)}-\frac{\rho\cdot\boldsymbol{b}_{(1)}}{\rho\cdot v_{2}}v_{2}^{\mu}.
\end{align}
Then, the quantity which must be differentiated is
\begin{align}
    |b|^{2}=-\boldsymbol{b}_{(1)}^{2}-\frac{(\rho\cdot\boldsymbol{b}_{(1)})^{2}}{(v_{2}\cdot\rho)^{2}}.
\end{align}
Its derivative is
\begin{align}
	\frac{\partial}{\partial \boldsymbol{b}_{{(1)},\mu}} |b|^2 = - 2 \boldsymbol{b}_{(1)}^{\mu}  - 2 \frac{\rho\cdot \boldsymbol{b}_{(1)} }{(\rho\cdot v_2)^2} \Pi_{{\rm 3d},(1)}^{\mu\nu}\rho_{\nu} .
\end{align}
This result is orthogonal to both $\rho^{\mu}$ and $v_{2}^{\mu}$, and holds with or without spin dependence.

Accounting for the Compton-amplitude contact terms further requires that we evaluate the integrals in \cref{eq:WaveformContactIntegral}:
\begin{align}
    \frac{\mathcal{J}^{\mu_{1}\dots\mu_{n}}_{(1)}(b_{(1)})}{v_{1}\cdot\rho}&=I_{w}^{\mu_{1}\dots\mu_{n}}[\{x,b_{1},b_{2}+i\mathfrak{a}_{2}\};1/q_{2}^{2}].
\end{align}
Computing with the method in appendix C of ref.~\cite{Cristofoli:2021vyo}, the rank-0 integral is
\begin{align}
    \mathcal{J}_{(1)}(b_{(1)})&=-\frac{1}{4\pi|\boldsymbol{b}_{(1)}|},
\end{align}
where $\boldsymbol{b}_{(1)}^{\mu}=b_{(1)\nu}\Pi_{{\rm 3d,}(1)}^{\mu\nu}$.
Differentiating,
\begin{align}\label{eq:ContactIntRank1}
    \mathcal{J}_{(1)}^{\mu}(b_{(1)})&=\frac{-i\partial}{\partial\boldsymbol{b}_{(1),\mu}}\mathcal{J}_{(1)}(b_{(1)})=\frac{i}{4\pi}\frac{\boldsymbol{b}^{\mu}_{(1)}}{|\boldsymbol{b}_{(1)}|^{3}}.
\end{align}
We have corroborated this expression through explicit computation as for the rank-0 case, thus verifying the validity of this derivative operation even in the presence of spin.
\Cref{eq:ContactIntRank1} is in agreement with ref.~\cite{Jakobsen:2021lvp}, only the result here encodes effects at all spin orders.
Generating higher ranks is straightforward with the integral written in this form, keeping in mind that
\begin{align}
    \frac{\partial\boldsymbol{b}_{(1)}^{\mu}}{\partial\boldsymbol{b}_{(1),\nu}}=\Pi_{{\rm 3d,}(1)}^{\mu\nu},
\end{align}
such that the result remains orthogonal to $v_{2}^{\mu}$.

\subsection{Memory-effect integration}

For the leading-order memory effect, the only integral we need is \cite{Kosower:2018adc}
\begin{align}
    I_{m}^{\mu}[b;1/q^{2}]&=-\frac{i}{2\pi\sqrt{\gamma^{2}-1}}\frac{b^{\mu}}{b^{2}},
\end{align}
where $|x|\equiv\sqrt{-x^{2}}$.
In terms of this, the all-spin leading-order memory effect is obtained by a redefinition of the impact parameter.

At next-to-leading order, all integrals we need can be obtained by differentiating the base integral \cite{Aoude:2020ygw}
\begin{align}
    I_{m}[b;1/|q|]&=\frac{1}{2\pi\sqrt{\gamma^{2}-1}}\frac{1}{|b|}.
\end{align}
In terms of this, the rank-$n$ integral is
\begin{align}
    I^{\mu_{1}\dots\mu_{n}}_{m}&=-i\partial_{b}^{\mu_{n}}I^{\mu_{1}\dots\mu_{n-1}}_{m}.
\end{align}
When employing this relation, it is crucial to keep in mind that
\begin{align}
    \frac{\partial b^{\mu}}{\partial b_{\nu}}&=\Pi^{\mu\nu}_{{\rm 2d}},
\end{align}
such that the result of the differentiation remains in the 2-plane containing $b^{\mu}$.
We needed an arbitrary number of such derivatives to present all-order-in-spin results for the anti-aligned configuration.
This task was simplified in two ways.
First, the complexity of the derivatives is reduced by considering instead $I_{m}[q^{2n}/|q|]$, which enter in the calculation for $0\leq n\leq3$.
Then, taking advantage of the fact that $v_{i}\cdot\mathfrak{a}_{\rm aa}=b\cdot\mathfrak{a}_{\rm aa}=0$, we were able to find a closed form for the projection of the rank-$2k$ Fourier transform into the hyperplane orthogonal to $\mathfrak{a}_{\rm aa}^{\mu_{i}}\mathfrak{a}_{\rm aa}^{\nu_{i}}$ for $1\leq i\leq k$.
Specifically,
\begin{align}
    &\prod_{i=1}^{k}\left(\mathfrak{a}_{{\rm aa},\mu_{i}}\mathfrak{a}_{{\rm aa},\nu_{i}}-\eta_{\mu_{i}\nu_{i}}\mathfrak{a}_{{\rm aa}}\cdot\mathfrak{a}_{{\rm aa}}\right)I^{\mu\mu_{1}\nu_{1}\dots\mu_{k}\nu_{k}}_{m}[b;q^{2n}/|q|]\notag \\
    &=-i\frac{4^{n}(1/2)_{n}}{(1)_{n}}\frac{(-4)^{k}(3/2)_{k+n}(1)_{k+n}}{2\pi\sqrt{\gamma^{2}-1}}\frac{\mathfrak{a}_{\rm aa}^{2k}b^{\mu}}{|b|^{2k+2n+3}},
    \label{eq:NLOMemoryAAFTRank1}\\
    &\prod_{i=1}^{k}\left(\mathfrak{a}_{{\rm aa},\mu_{i}}\mathfrak{a}_{{\rm aa},\nu_{i}}-\eta_{\mu_{i}\nu_{i}}\mathfrak{a}_{{\rm aa}}\cdot\mathfrak{a}_{{\rm aa}}\right)I^{\mu\nu\mu_{1}\nu_{1}\dots\mu_{k}\nu_{k}}_{m}[b;q^{2n}/|q|]\notag \\
    &=-\frac{4^{n}(1/2)_{n}}{(1)_{n}}\frac{(-4)^{k}(3/2)_{k+n}(1)_{k+n}}{2\pi\sqrt{\gamma^{2}-1}}\frac{\mathfrak{a}_{\rm aa}^{2k-2}}{|b|^{2k+2n+3}}\notag \\
    &\times\left[\mathfrak{a}_{\rm aa}^{2}\Pi^{\mu\nu}+(2k+2n+3)\mathfrak{a}_{\rm aa}^{2}\frac{b^{\mu}b^{\nu}}{|b|^{2}}+\frac{k}{n+1}\mathfrak{a}_{\rm aa}^{\mu}\mathfrak{a}_{\rm aa}^{\nu}\right].\label{eq:NLOMemoryAAFTRank2}
\end{align}
These integrals are sufficient for completely determining the next-to-leading-order memory effect in the anti-aligned configuration to all spin orders, including in the presence of higher-spin-order Compton-amplitude contact terms than those considered here.

For the case of more generally oriented spins, we evaluated the next-to-leading-order memory effect up to sixth order in spin.
This necessitated up to six derivatives of \cref{eq:NLOMemoryAAFTRank1,eq:NLOMemoryAAFTRank2} with $k=0$.

\bibliography{BibliographyWaveform.bib}

\end{document}